\documentclass{ws-mplb}

\usepackage{cancel}

\newcommand{\bn}{\begin{enumerate}}
\newcommand{\en}{\end{enumerate}}
\newcommand{\ba}{\begin{eqnarray}}
\newcommand{\ea}{\end{eqnarray}}

\usepackage{graphicx,color}

\newcommand{\be}{\begin{equation}}
\newcommand{\ee}{\end{equation}}

\newcommand{\et}{{\it et al. }}
\newcommand{\ete}{{\it et al.}}

\def\prl{{ Phys. Rev. Lett. }}

\newcommand{\gd}[3]{Gd$_{#1}$Fe$_{#2}$Co$_{#3}$}

\newcommand{\tb}[2]{Tb$_{#1}$Co$_{#2}$}
\newcommand{\tbfe}[2]{Tb$_{#1}$Fe$_{#2}$}

\newcommand{\tbfeco}[3]{Tb$_{#1}$Fe$_{#2}$Co$_{#3}$}

\newenvironment{mybox}[2]
{ \fbox{
\parbox{#1cm}
{
#2
}
}
}

\begin{document}

\markboth{Zhang, Murakami, Si, Bai, and George }{Understanding
  all-optical spin switching}

%
\catchline{}{}{}{}{}
%

\title{Understanding all-optical spin switching:\\ Comparison between
  experiment and theory
}

\author{G. P. Zhang$^*$ and  M. Murakami
}

\address{Department of Physics, Indiana State University,
   Terre Haute, IN 47809, USA
$^*$gpzhang.physics@gmail.com}

\author{M. S. Si
}

\address{Key Lab for Magnetism and Magnetic
  Materials of the Ministry of Education, Lanzhou University, Lanzhou
  730000, China
}

\author{Y. H. Bai}

\address{Office of Information Technology, Indiana State
  University, Terre Haute, IN 47809, USA}

\author{Thomas F. George}

\address{Office of the Chancellor, 
  Departments of Chemistry \& Biochemistry and Physics \& Astronomy
  University of Missouri-St. Louis, St.  Louis, MO 63121, USA }

\maketitle

\begin{history}
\received{(\today)}
\revised{(Day Month Year)}
\end{history}

\begin{abstract}
{Information technology depends on how one can control and manipulate
  signals accurately and quickly. Transistors are at the core of
  modern technology and are based on electron charges. But as the
  device dimension shrinks, heating becomes a major problem.  The
  spintronics explores the spin degree of electrons and thus bypasses
  the heat, at least in principle.  For this reason, spin-based
  technology offers a possible solution.  In this review, we survey
  some of latest developments in all-optical switching (AOS), where
  ultrafast laser pulses are able to reverse spins from one direction
  to the other deterministically. But AOS only occurs in a special
  group of magnetic samples and within a narrow window of laser
  parameters. Some samples need multiple pulses to switch spins, while
  others need a single-shot pulse. To this end, there are several
  models available, but the underlying mechanism is still under
  debate. This review is different from other prior reviews in two
  aspects. First, we sacrifice the completeness of reviewing existing
  studies, while focusing on a limited set of experimental results
  that are highly reproducible in different labs and provide actual
  switched magnetic domain images. Second, we extract the common
  features from existing experiments that are critical to AOS, without
  favoring a particular switching mechanism. We emphasize that given
  the limited experimental data, it is really premature to identify a
  unified mechanism. We compare these features with our own model
  prediction, without resorting to a phenomenological scheme. We hope
  that this review serves the broad readership well.  }
\end{abstract}


\section{Introduction}




Computing technology demands high-speed operation and miniaturization
of computing bits.\cite{theis2010a} It is remarkable that the computer
clock frequency has improved steadily for the last decade, but then
has plateaued around 2005, because the simple constant-electric field
scaling rules break down.\cite{theis2010b} Both technical constraints
(physics) and cost constraints (economics) are responsible for the
clock frequency plateau.\cite{theis2010b} In the constant-field
scaling, the threshold voltage must be reduced as the clock frequency
increases, but this very reduction results in increase in OFF current,
which is unacceptable to maintain a high ON/OFF current ratio. The
second problem is associated with the gate insulator thickness
reduction, which increases gate current leakage. A way out of this is
to use the constant voltage scaling. But it comes with an huge
increase in areal power density. So this is
unsustainable.\cite{theis2010b} To meet the insatiable demand on
computing technology, one must pursue other means.

\begin{figure}
\centerline{\psfig{file=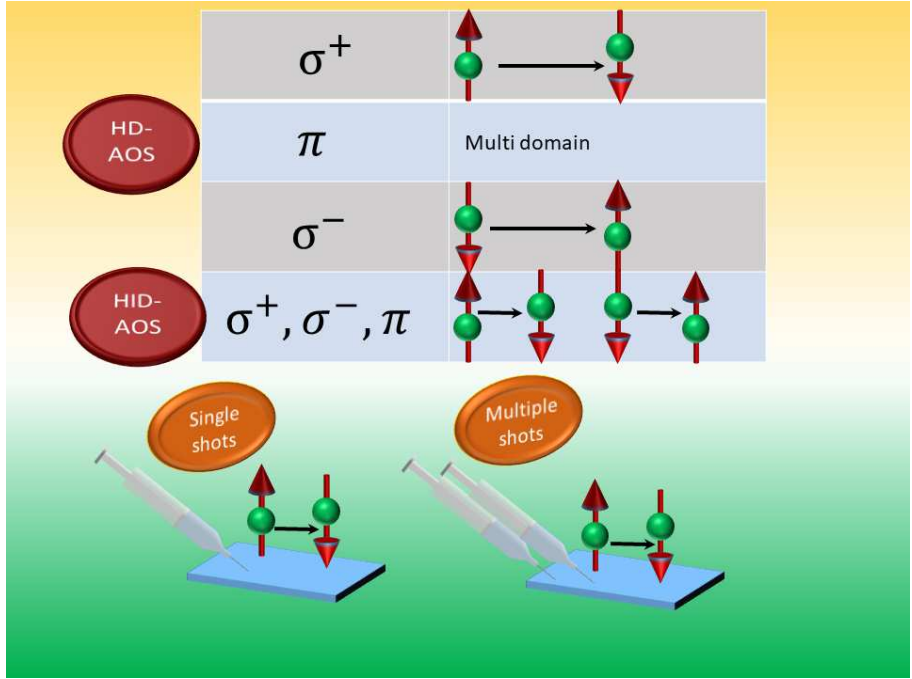,width=12cm,angle=0}}
\caption{Overview of all-optical spin switching. Based on the laser
  helicity, one can distinguish two major kinds of AOS: One is
  helicity-dependent and the other is helicity-independent. Based on
  the number of pulses used in switching, there are single-shot and
  multiple-shot switchings.  }
\label{fig1}
\end{figure}

All-optical spin switching (AOS)\cite{stanciu2007} combines the
speed\cite{eric} that an ultrafast laser pulse delivers and the
powerful storage capability that existing magnetic media offer, yet
free of a magnetic field. It is still at the early stage of
development, but results are promising. Figure \ref{fig1}
schematically shows several possible switching channels. In general,
AOS can be classified into two broad categories, depending on how the
helicity affects AOS and how many laser pulses are needed to switch
spins, respectively. The first category refers to the
helicity-dependent all-optical switching, HD-AOS, where the
right-circularly polarized light ($\sigma^+$) switches a spin from up
to down, while the left-circularly polarized light ($\sigma^-$) from
down to up. The linearly polarized light ($\pi$) only creates
multidomains, with mixed up and down spins. By contrast, in the
helicity-independent all-optical switching, HID-AOS, $\sigma^+$,
$\sigma^+$ and $\pi$ all switch spins. This is in contrast to ordinary
thermomagnetic switching due to local laser heating and dipolar
interactions, where it only leads to a one-time reversal without
switching back the spins.\cite{hansen1988}

Besides the above elementary operations, AOS is found highly laser-
and material-specific, where often one finds that spins in the same
type of materials are both switched optically and electrically.  A
large fraction of AOS materials require multiple pulses to switch
spins. So far, only four types of materials --  GdFeCo,\cite{ostler2012}
Pt/Co/Gd,\cite{lalieu2017} Co/Pt/Co/GdFeCo,\cite{gorchon2017} and
Pt/Co/Pt\cite{vomir2017} -- allow a single shot switching (see
Fig. \ref{fig1}).  They are among the most promising materials to be
used in the future, and they  can create tunable topological magnetic
structures such as Skyrmions.\cite{finazzi2013}  

 There are several excellent review articles available.  Kirilyuk and
 coworkers\cite{rasingreview} reviewed the status of the field before
 2010, which is quite comprehensive and includes other branches of
 femtomagnetism.\cite{ourreview} Two newer
 reviews\cite{mplb16,elhadri2017} cover major parts of prior research
 in this field.  Our current review does not aim to be comprehensive,
 so we regret that many excellent references can not be cited. The
 review instead complements the prior reviews by focusing on highly
 reproducible experiments where magnetic domain images are taken.  As
 research is ongoing,\cite{lu2018} we consider our review as an
 alternative to the existing understanding of AOS.


The rest of the paper is arranged as follows. In Sec. II, we briefly
review the history of AOS\cite{stanciu2007} but focus on those
specific materials that allow AOS. These materials are provided in a
table for the reader, with hope to find a common feature among all AOS
materials. Section III is devoted to the role of the spin moment and
spin configuration in AOS, where we discuss how materials do not
switch, impact of reduced dimensionality, a practical method of
extraction of spin angular momentum from experimental data, and
dynamical simulation. In Sec. IV, we present a simple theory for
all-optical switching, where we connect the inverse Faraday theory
with our model and first-principles results. Section V highlights the
significance of orbital angular momentum on perpendicular magnetic
anisotropy and AOS. Section VI is based on the latest experimental
results in single-shot AOS ferromagnets. Finally, we conclude this
paper in Sec. VII.

\section{Discovery of AOS}

In 2007, Stanciu and coworkers\cite{stanciu2007} discovered that
exposing a ferrimagnet $\rm Gd_{22}Fe_{74.6}Co_{3.4}$ to a 40-fs
800-nm laser pulse creates a permanent spin reversal.  The images of
magnetic domains, before and after laser excitation, record the
remarkable switching. The right-circularly polarized light
($\sigma^+$) switches the down domain to the up domain. If the
magnetic domain is already up, then there is no effect on the domain.
The left-circularly polarized light ($\sigma^-$) switches the up
domain to down domain and has no effect on the down domain. If a domain
is exposed to linearly polarized light ($\pi$), it does not switch,
but instead breaks the original domain into smaller domains randomly
oriented up or down.  Its broad appeal to both materials scientists,
and optical and electric engineers is almost immediate and results in
plenty of experimental investigations.

For a long time, ferrimagnetic rare-earth transition metal
alloys\cite{stanciu2007,liu2015} have remained the only material
showing AOS. This led some researchers to speculate the crucial role
of antiferromagnetic orders between the rare-earth and
transition-metal sublattices.  By contrast, many antiferromagnetic
materials, such as TmFeO$_3$\cite{kimel2004} and
DyFeO$_3$,\cite{kimel2005} which have a ``correct'' coupling, do not
switch their spins permanently; instead, they only reorient their
spins.  To get a glimpse of the hot debates among different research
groups, in Table \ref{tab1} we list 31 most intensively investigated
AOS compounds, together with 4 non-AOS compounds (first four
entries). We also list magnetic orderings, underlying mechanisms
proposed or disproved (the entries with slanted lines), and whether
the switching is HD-AOS or HID-AOS. Unfortunately, not all the studies
have discussed the underlying mechanism, so we leave them blank. In
some cases, several mechanisms are proposed, but we only choose one or
two. The acronyms are explained in the caption of the table.  These 4
non-AOS compounds are used as a counter-example that antiferromagnetic
ordering and the inverse-Faraday effect may not be enough for AOS. All
31 AOS compounds have actual magnetic domain images taken before and
after laser exposure. We specifically avoid materials that only one or
two groups are familiar with.

\begin{table}
\caption{Chronicle of AOS materials whose magnetic domain images are
  taken.  Ordering refers to magnetic ordering.  IFE: inverse Faraday
  effect. SF-SRS: spin-flip stimulated Raman scattering.  FIM:
  ferrimagnetic. AFM: antiferromagnetic. PM: paramagnetic.  Under
  Mechanism, only a selected few are listed.  Linear: linear reversal;
  HD-AOS: helicity-dependent all-optical switching; HID-AOS:
  helicity-independent all-optical switching; T$_{\rm comp}$:
  compensation temperature dependent; MCD: magnetic circular
  dichroism; SDC: superdiffusive current; LR: low remanence; DM:
  magnetic domain size; ST: stochastic. These underlined compounds are
  the only ferromagnets that show a single-shot switching. The slanted
lines denote those that are disapproved by a referenced paper.}
\begin{tabular}{lccccc}
\hline
\hline
Compound& Ordering  & Mechanism & AOS/Non-AOS  & Ref.\\
\hline
TmFeO$_3$ & AFM                  &        & Non-AOS          & \cite{kimel2004}\\
DyFeO$_3$ & AFM           & IFE & Non-AOS          & \cite{kimel2005}  \\
HoFeO$_3$ &AFM                   & IFE  & Non-AOS          & \cite{kimel2009}  \\ 
NaTb(WO$_4$)$_2$ &PM             & IFE  & Non-AOS          &\cite{jin2010} \\
\gd{22}{74.6}{3.4}&FIM           &IFE   & HD-AOS           &\cite{stanciu2007}\\
\gd{22}{74.6}{3.4}&FIM           & \cancel{Thermal}   & AOS              &\cite{hohlfeld2009}\\
\gd{24}{66.5}{9.5}&FIM           & Linear  &HD-AOS               &\cite{vahaplar2009}\\
\gd{26}{64.7}{9.3}&FIM           &\cancel{IFE/SF-SRS}       &HD-AOS               &\cite{steil2011}\\
\gd{23}{68}{9}    &FIM           &IFE                       &HD-AOS               &\cite{ohkochi2012}\\
\gd{x=20\leftrightarrow28}{90-x}{10} &FIM &IFE/Linear       &H(I)D-AOS            &\cite{vahaplar2012}\\
\gd{26}{65}{9} &FIM              &MCD                       &HD-AOS &\cite{khorsand2012}\\
\gd{24,25}{66.5}{9.5} &FIM&Thermal&HID-AOS&\cite{ostler2012}\\
\gd{24,25}{65.6}{9.4} &FIM&Thermal                       &HID-AOS&\cite{ostler2012}\\
\gd{24}{66.5}{9.5}&FIM           &                          &H(I)D-AOS&\cite{alebrand2012a}\\   
\tb{x=0.12\leftrightarrow0.34}{1-x}&FIM        & T$_{\rm comp}$            &H(I)D-AOS               &\cite{alebrand2012}\\
\tbfe{x=19\leftrightarrow38.5}{100-x} &FIM     & \cancel{T$_{\rm    comp}$}&HD-AOS&\cite{hassdenteufel2013}\\    
Co/Ir/CoNiPtCo/Ir,\tb{26}{74}&FIM&T$_{\rm comp}$            &HD-AOS & \cite{mangin2014}\\
Tb/Co multilayer&FIM& T$_{\rm comp}$   &HD-AOS & \cite{mangin2014}\\
\tbfe{36}{64}/\tbfe{19}{81} & FIM & LR &HD-AOS & \cite{schubert2014a}\\
\tbfe{29}{71},\tbfe{34}{66} &FIM  & LR &HD-AOS &\cite{hassdenteufel2015}\\
\tbfe{30}{70} &FIM &conductivity &HD-AOS &\cite{hassdenteufel2014}\\
\tbfeco{22}{69}{9} &FIM &IFE,\cancel{MCD} &HD-AOS &\cite{gierster2015}\\
\tb{x=8\rightarrow 14.5}{100-x}($<6.5$nm)&FIM& DM, \cancel{LR}  &HD-AOS & \cite{elhadri2016a}\\
\tb{x=16.5\rightarrow
  30.5}{100-x}($<15$nm)&FIM& DM, \cancel{LR} &HD-AOS & \cite{elhadri2016a}\\
\tbfe{x=22\rightarrow
  34}{100-x}(5-85nm)&FIM& {LR} &HD-AOS & \cite{hebler2016}\\
\underline{$\rm Pt/Co/Gd$}
&FIM& Thermal  &HID-AOS & \cite{lalieu2017}\\
$\rm [Co(4\AA)/Pt(7\AA)]_{2\rightarrow 3}$ &FM& &HD-AOS 
&\cite{lambert2014}\\
$\rm Pt/Co(6\AA\leftrightarrow 15\AA)/Pt$ &FM& &HD-AOS 
&\cite{lambert2014}\\
$\rm [Pt/Co_{1\it -x}Ni_{\it x}(6\AA)]_{2\rightarrow 4}$ &FM& &HD-AOS 
&\cite{lambert2014}\\
$\rm Cu/[Ni(5\AA)/Co(1\AA)]_{2}/Ni/Cu$ &FM&\cancel{SDC} &HD-AOS 
&\cite{lambert2014}\\
$[\rm Co(2\AA)/Ni(6\AA)]_2$
&FM& DM &HD-AOS & \cite{elhadri2016a}\\
$[\rm Pt(7\AA)/Co(6\AA)]_{1-2}$
&FM& DM &HD-AOS & \cite{elhadri2016a}\\
$\rm FePt$
&FM& ST &HD-AOS & \cite{john2017}\\
\underline{$\rm Co/Pt/Co/GdFeCo$}
&FIM/FM& \cancel{transport} &HID-AOS & \cite{gorchon2017}\\
$\rm [Co/Pt]/Cu/GdFeCo$
&FIM/FM& transport &HID-AOS & \cite{iihama2018}\\
\underline{$\rm Pt/Co/Pt$}
&FM& \cancel{IFE} &HID-AOS & \cite{vomir2017}\\
\hline\hline
\label{tab1}
\end{tabular}
\end{table}

This table is particularly important. Several common themes appear and
can be summarized as follows:

\vspace{0.5cm}

\mybox{11}
{\sf
\bn
\item  Both ferrimagnetic
and ferromagnetic compounds allow AOS.

\item All the materials tend to be very thin.

\item Nearly all the materials,\cite{guyader2015} except one
  sample,\cite{ostler2012} have perpendicular magnetic anisotropy.

\item The list is dominated by rare-earth $4f$ compounds and $5d$
  materials. Therefore, any theory that aims to explain AOS must start
  from the above three basic features.

\item There is no agreement on a unified mechanism. Proposed
  mechanisms tend to eliminate each other out. Any new mechanism must
  be conceived at least one level more microscopic and more
  fundamental than the existing ones.

\item What is not shown in the table is that the laser intensity has a
  narrow region to switch spins, beyond which AOS does not occur. The
  laser intensity is lower than that for demagnetization.

\en
}


\section{Role of Spin Angular Momentum and Spin Configuration in AOS}

\subsection{Materials that do not switch -- Iron nanoarrays}

\begin{figure}
\centerline{\psfig{file=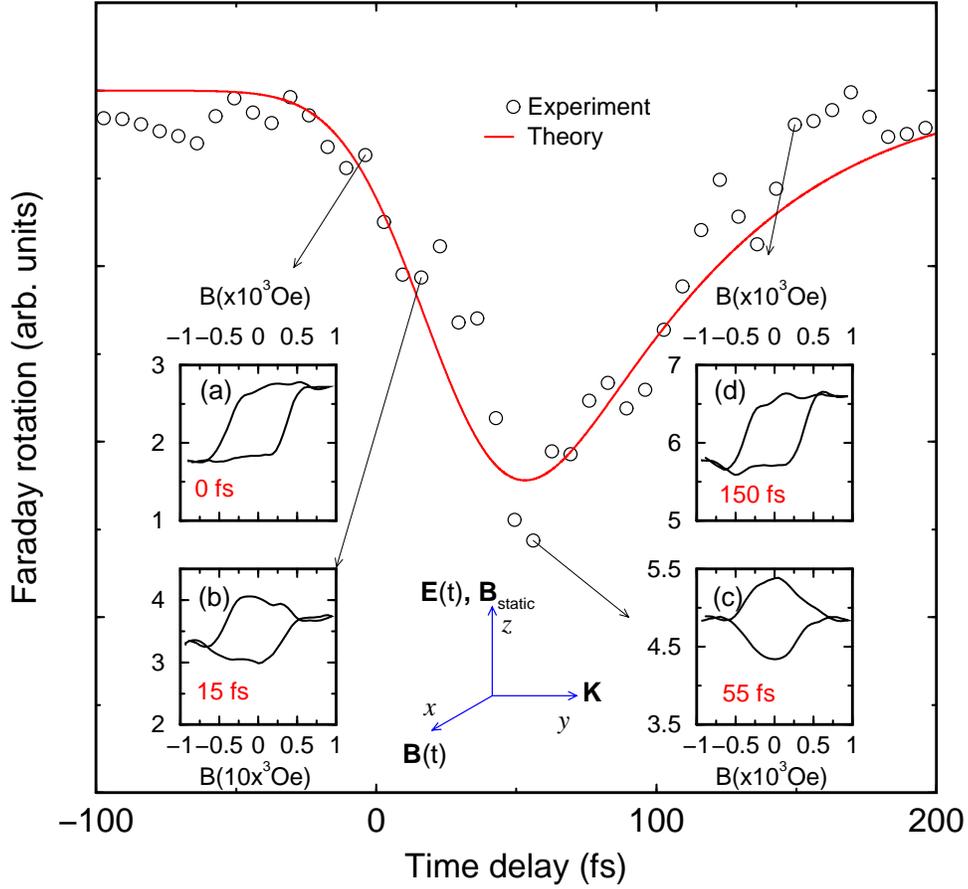,width=12cm,angle=270}}
\caption{Experimental Faraday rotational angle in Fe nanoarrays as a
  function of time delay between the pump and probe (empty circles)
  and theoretical spin change with time (solid line). (a)-(d) show the
  hysteresis loops at delays of 0, 15, 55 and 150 fs,
  respectively. The pump fluence is fixed at $\sim$ 25
  $\mu$J/cm$^2$. The loops were recorded by the probe beam with a
  photoelastic modulator (PEM) frequency. Before the arrival of pump
  pulses, a simple square shaped loop is observed, as expected for the
  easy axis. After the pump pulse excitation, the magnetization vector
  rotates away from the longitudinal direction, and the hysteresis
  loop reaches a full diamond shape at 55 fs. The loop then retrieves
  and recovers to the original square shape at 150 fs.  Inset:
  Configurations of the static magnetic field ${\bf B}_{\rm static}$
  and laser electric ${\bf E}(t) $ and magnetic fields ${\bf
    B}(t)$. The laser propagates along the $y$ axis with wavevector
  ${\bf K}$.  Reproduced from Y. Ren, W. Lai, Z. Cevher, Y. Gong and
  G. P. Zhang, Applied Physics Letters {\bf 110}, 082404 (2017), with
  the permission of AIP Publishing.\protect\cite{ren2017} }
\label{fig2}
\end{figure}

There are many more magnetic materials which do not switch their spins
under laser excitation than those which do switch. Most ferromagnets
only demagnetize. To understand why they do not switch, we present an
example first.  In 2017, Ren \ete\cite{ren2017} employed a group of
Fe, Fe/Pt, and Fe$_3$O$_4$ nanoarrays, with thickness of 50 nm to 200
nm. These nanoparticles have a diameter of about 50 nm, and the
center-to-center distance is 100 nm.  They shined a 50-fs and 800-nm
laser pulse on to the samples.  They found that upon laser excitation,
the field-free Faraday rotation angle in Fe nanoarrays is sharply
reduced, (see Fig. \ref{fig2}). It may appear that the sample simply
demagnetizes. However, this is not the whole story. At 0 fs, the
hysteresis loop has a normal rectangular shape, but around 15 fs, the
signal at zero field is stronger than at nonzero field. This indicates
that the spin, under joint effects of the laser field and applied
magnetic field, deviates from its original direction.  One sees that a
diamond shape is completely formed around 55 fs (see Fig. \ref{fig2}),
regardless of whether they employ $\sigma^+$, $\sigma^+$, or $\pi$
pulses.\cite{ren2017} After 150 fs, the normal rectangular shape is
restored.  This shows that besides the demagnetization, the spin also
cants out of the sample surface, but spin switching is not observed.

In the experiment, the pump incident to the sample has fluence of
25$\rm \mu J/cm^2$. The probe pulse is incident at 35$^\circ$ with
respect to the sample normal. The pump fluence has to be kept low to
cant spins; if the pump fluence is higher, it only demagnetizes the
sample without spin canting. This shows that spin canting needs much
less energy.  To explain this spin canting, we employ a simple theory
restricted to a single site as outlined previously,\cite{epl15} which
will be explained in detail below.  The model includes the kinetic and
potential energy terms, both of which are expressed in real
spaces. This allows us to compute the expectation values of the
electron velocity and position. The laser field is included through a
dipole term. The model is similar to the classical harmonic oscillator
model. But what is different from the traditional magneto-optics
formalism, is that we include a spin-orbit coupling in real space,
instead of a magnetic field.\cite{epl15} This surprisingly captures
the early switching of the spin. It was later found that the
spin-orbit torque plays a role in AOS.\cite{epl16}

The simulation uses the same laser parameters as the experimental
ones, and has a single fitting parameter for the recovery of the
spin. The static magnetic field and laser field directions are shown
in the inset of Fig. \ref{fig2}.  The solid line shown in
Fig. \ref{fig2} is our theoretical result. One can see that the theory
matches the experimental results quite well. The theory shows that the
spin cants out of plane and oscillates within the $xy$
plane.\cite{ren2017} This is the first experimental verification of
the model,\cite{epl15} which is beyond the semiempirical
model.\cite{ostler2012} Our model further shows that regardless of
spin configuration, the spin switching is not observed for in-plane
magnetic anisotropy in this system. The reason is because the in-plane
anisotropy is unfavorable to AOS and the spin angular momentum is too
small. \cite{jpcm17b}

In fact, AOS was never found in any of three $3d$ element
ferromagnets, Ni, Fe and Co, and $4f$ rare earth magnets, Gd and
Tb. In these materials, even spin canting by light is rare, though
canting can be induced magnetically.\cite{patrick2018} Ren's study
reveals an important fact that the dimensionality matters. In Ni thin
films, only demagnetization is observed.\cite{eric} Another piece of
information from Ren's work is that those nanoarrays all have in-plane
magnetic anisotropy, in contrast to all the compounds listed in Table
\ref{tab1} which have perpendicular magnetic anisotropy (PMA).  So,
the next question is what the reduced dimensionality and PMA imply
microscopically.

\subsection{Impact of thickness and reduced dimension}

\begin{figure}
\centerline{\psfig{file=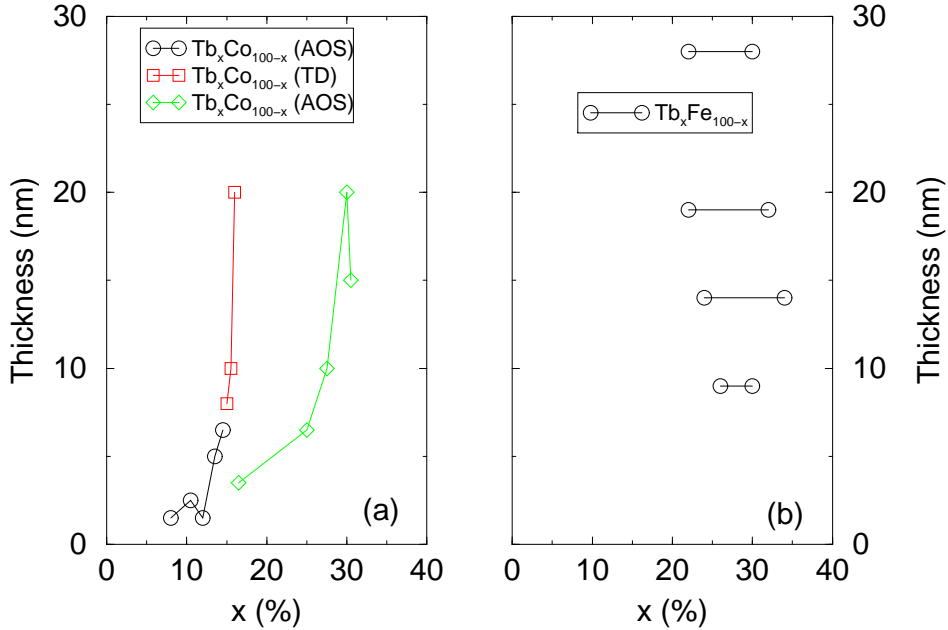,width=0.7\columnwidth,angle=270}}
\caption{Relation between thickness and composition in AOS
  materials. (a) TbCo-based AOS materials are normally very thin (see
  circles). If the thickness increases, the Tb composition $x$ must
  increase (see diamonds). Otherwise, thermal demagnetization (TD) is
  observed (see squares). This shows the importance of the large spin
  moment of each constituent. The original data are
  from Ref. 46.
(b) TbFe-based AOS alloys are also
  thin. But their thickness can be slightly larger than those in (a),
  since Fe has a larger spin moment than Co. The original data are
  from Ref. 36.}
\label{compare}
\end{figure}

 One obvious answer is that the reduced dimensionality strengthens the
 perpendicular magnetic anisotropy. But this can not be the only
 reason for AOS, since many magnetic thin films have PMA and many do
 not switch spins under laser excitation. We choose two sets of
 experimental data from two independent groups. One is from El Hadri
 \ete\cite{elhadri2016} and the other is from Hebler
 \ete\cite{hebler2016} They used two different samples, \tb{x}{100-x}
 and \tbfe{x}{100-x}, respectively, with various thicknesses. We
 caution that when the thickness is thin, the nominal thickness may
 differ from the actual thickness.  We see the thickness as a good
 gauge to test the effect of the dimensionality on AOS, so we plot the
 thin film thickness with the composition, where AOS or thermal
 demagnetization (TD) is observed. Figure \ref{compare}(a) shows that
 AOS occurs in very thin films. If the composition of Tb ($x$) is
 smaller, the window of thickness for AOS is narrower. As $x$
 increases, the window of thickness gets wider. The AOS-allowable
 thickness reaches 20 nm if $x$ increases to 30\%. $x$ can not
 increase forever, since the composition of Co becomes too low to
 switch spins. This is probably the reason behind the peak in
 Fig. \ref{compare}(a).  A direct consequence of $x$ is that it
 affects the effective spin moment for each element. If this is true,
 we should expect that \tbfe{x}{100-x} allows a thicker sample for
 AOS.  Figure \ref{compare}(b) shows that it is indeed true that for a
 similar composition $x$ around 30\% in \tbfe{x}{100-x}, the maximum
 thickness reaches 28 nm, while in \tb{x}{100-x}, it is only 20
 nm. According to Albrecht,\cite{albrecht2018} for a Tb content of
 about 28\%, AOS is even possible for TbFe films at least as thick as
 85 nm, the maximal thickness investigated.  These two results seem to
 suggest that there is a minimum spin angular momentum which each
 element in these two compounds has to exceed before AOS can
 occur. The reduced thickness facilitates a boost of the spin
 moment. However, these two experimental results are not enough to
 make a definitive conclusion. Naturally, reduced dimensionality also
 affects other properties. For instance, Yuan \et \cite{yuan2003}
 showed that in Co/Pt multilayers the demagnetization field inversely
 depends on cobalt thickness.

\subsection{Extracting spin angular momenta from real materials}

Let us to find how large the spin angular momenta are in real
samples. To extract spin moments from each element of a complex
amorphous sample is highly nontrivial.  Nearly all the experiments
give magnetization, not spin moment. A change from magnetization to
spin moment needs the volume of a sample, which is rarely given. Added
to the complication is magnetization in the unit of $\rm kA/m$ or
emu/cc (electromagnetic unit cubic centimeter).

In 2016, we found a method\cite{epl16} which is not perfect but works
reasonably well. However, the scheme was never published formally, but
it was posted on Arxiv.org.\cite{arx2016} Interestingly, a similar
method was used by others.\cite{gridnev2018} The following material is
adopted from the post with minor changes.  What we do is to replace
the sample volume by the unit cell volume. The unit cell of an
amorphous sample is approximated by the unit cells of element
constituents. Then, we use the same proportional relation to compute
the spin moments of each constituent. To show how this works, we
consider a rare-earth-transition metal alloy, $R_xT_{1-x}$, where R
stands for Tb or Gd, and T stands for Fe. We ignore Co since its
concentration is too low.  We first compute the effective volume \be
V_{eff}=xV_{\rm R}+(1-x)V_{\rm T}, \ee where $V_{\rm R}$ and $V_{\rm
  T}$ are the supercell volumes of the pure elements R and T,
respectively.  Tb has a hcp structure, with the lattice constants
$a=3.601\rm \AA$ and $c=5.6936\rm \AA$; Fe has a bcc structure with
$a=2.8665\rm \AA$.  Then we multiply the magnetization $m$ for a
particular element by $V_{eff}$ to get the effective spin moment for
the alloy, i.e., $M_{eff}= mV_{eff}$.  Since $M_{eff}$ is in the units
of [Am$^2$], we convert it to the Bohr magneton $\mu_B$, with the
conversion factor of $0.10783 \times 10^{-3}$ (i. e.,
$10^{-3}/9.274$).

Szpunar and Kozarzewski \cite{szpunar1977} carried out extensive
calculations on transition-metal and rare-earth intermetallic
compounds by comparing their results with the experimental ones, and
concluded that it is reasonable to assume that the average magnetic
moments of the transition metals and of the rare earth metals are
roughly independent of structures. Hansen and Witter\cite{hansen1988}
specifically tested the linear relationship between the uniaxial
anisotropy constant $K_u$ measured by the torque method and the Tb
content and found the single-ion contribution of the terbium atom, so
the TbFe alloy has $K_u({\rm Temp})=x K_u^{\rm Fe}({\rm
  Temp})+(1-x)K_u^{\rm Tb}({\rm Temp})$, where ${\rm Temp}$ is
temperature.  In the same spirit, we approximate the effective spin
moment $M_{eff}$ as \be M_{eff}=xM_{\rm R}+(1-x)M_{\rm T}\equiv
M^{eff}_{\rm R} + M^{eff}_{\rm T}, \ee where $M_{\rm R}$ and $M_{\rm
  T}$ are the spin moments of pure R and T, respectively.  Here the
last equation defines the effective spin moment for R and T.

However, this single equation is not enough to compute $M_{\rm R}$ and
$M_{\rm T}$ since there are two unknowns for a single equation.  The
trick is that we use two sets of compositions, $x_1$ and $x_2$, so we
have two equations, \ba M_{eff}^{(1)}=x_1M_{\rm R}+(1-x_1)M_{\rm
  T}\\ M_{eff}^{(2)}=x_2M_{\rm R}+(1-x_2)M_{\rm T}, \label{x12} \ea where
$M_{eff}^{(1)}=m_R^{(1)}V_{eff}^{(1)}$ and
$M_{eff}^{(2)}=M_R^{(2)}V_{eff}^{(2)}$.  Here again we rely on the
assumptions that $M_{\rm R}$ and $M_{\rm T}$ do not change much with
composition change from $x_1$ to $x_2$.  When we choose $x_1$ and
$x_2$, we are always careful whether $M_{\rm R}$ or $M_{\rm T}$
changes sign, since experimentally the reported values are the
absolute values. In addition, it is always better to choose those $x_1$
and $x_2$ which have the same sign of $M_{\rm R}$ and $M_{\rm
  T}$. Choosing several different pairs of $(x_1,x_2)$ is crucial for a
reliable result.  Solving the above two equations, we can find $M_{\rm
  R}$ and $M_{\rm T}$.

Before we compute the spin angular momentum, we check whether the
computed spin moments $M_{\rm R}$ and $M_{\rm T}$ (in the units of
$\mu_B$) are close to their respective values of each pure element,
i.e., $M_{\rm Gd}^{\circ}=7.63 \mu_B$,\cite{kurz2002}
 $M_{\rm
  Tb}^{\circ}=9.34 \mu_B$, and $M_{\rm Fe}^{\circ}=2.2 \mu_B$.  If a
computed spin moment ($M_{\rm R}$ and $M_{\rm T}$) is far off from
those spin moments, this indicates that either our method or the
experimental result is not reliable.  Once the spin moment passes this
test, we proceed to convert the spin moment to spin angular momentum.


Our method works better for Gd alloys than Tb alloys, since the former
has nearly zero orbital angular momentum but the latter has a nonzero
orbital angular momentum.  For Gd and Fe, the orbital momentum is
largely quenched.  Assuming that the Lande $g$-factor is 2, we divide
the spin moments $M_{\rm R}$ and $M_{\rm T}$ by 2 to get the spin
angular momenta $S_{\rm R}$ and $S_{\rm T}$ in the unit of $\hbar$. To
get the effective spin angular momentum, we multiply $S_{\rm R}$ and
$S_{\rm T}$ with $x$ and $1-x$, respectively, i.e., \ba S_{\rm
  R}^{eff}&=&x S_{\rm R}\\ S_{\rm T}^{eff}&=&(1-x) S_{\rm T}.  \ea It
is these two effective spin angular momenta to which we apply our
minimum spin angular momentum criterion (see below for details).  For
Tb, our results have an uncertainty since its orbital angular momentum
in its alloys is unknown, although its orbital angular momentum in
pure Tb metal is 3.03$\hbar$.  Table \ref{tab0} shows the orbital-free
spin angular momentum for 11 alloys, where we adopt a simple cubic
structure for Fe since it matches the experimental values
better. There are multiple rows of spin angular momenta for the same
materials because there are several possible pairs of $(x_1,x_2)$ for
Eq. (\ref{x12}) that one can choose. For instance, \tbfe{30}{70}, the
composition-weighted spin angular momentum of Tb ranges between
$2.1506\hbar$ and $1.4952\hbar$.  The variation seen in those momenta
is because we choose multiple pairs of TbFe. For instance, if we
choose \tbfe{30}{70} and \tbfe{29}{71} as a pair, we get a
composition-weighted momentum; if we choose \tbfe{30}{70} and
\tbfe{22}{78}, we get another momentum. In principle, if the magnetic
properties among different compositions are independent of
composition, one should find the same momentum, but this is not always
the case since these materials are amorphous; different patches of
samples may have different structural and magnetic properties.  In the
Appendix, we have provided our computer code and one example, so the
reader can directly use it for his/her own research. We note that in
some cases, our method can even test the accuracy of the original
experimental data. However, our method does not work well for TbCo,
partly because there are only two data points.\cite{hassdenteufel2015}

Now, we have a table of effective spin angular momenta to work
with. Table \ref{tab0} shows that each constituent has a sizable spin
within AOS-allowed $x$ that is above $0.8\hbar$ as shown next.

\begin{table}
\caption{Computed effective spin angular momenta for each element in
  GdFeCo and TbFe alloys. Multiple pairs of alloys are used to compute
  the effective spin angular momentum for several compounds to
  demonstrate the range of the change in the spin angular
  momentum. The sign convention of the spin angular momentum is that
  either Gd or Tb has a positive value, while Fe has a negative value.
  A simple cubic structure is adopted for Fe. Two underlined entries
  are two examples that are explained in the Appendix.  }
\begin{tabular}{lllcc}
\hline
Alloy &$S^{eff}_{\rm Gd}(\hbar)$ &$S^{eff}_{\rm Fe}(\hbar)$ &
$S^{eff}_{\rm Tb}(\hbar)$ (orb. free) &$S^{eff}_{\rm Fe}(\hbar)$
(orb. free) \\
\hline
Gd$_{28}$Fe$_{63}$Co$_{9}$          & 1.3414   &  -1.1691  & -- &  --    \\
Gd$_{26}$Fe$_{64.7}$Co$_{9.3 }$     &   1.2456  &  -1.2006  & -- &  --   \\
Gd$_{25}$Fe$_{65.6}$Co$_{9.4}$      &  1.1517 &     -1.1777   & -- &--  \\
Gd$_{24}$Fe$_{66.5}$Co$_{9.5 }$     &  1.2262   &  -1.3017  & -- &  --   \\
Gd$_{24}$Fe$_{66.5}$Co$_{9.5 }$     &  1.0867   &    -1.0113   & -- &  --   \\
 Gd$_{22}$Fe$_{68.2}$Co$_{9.8}$     &   1.1241  &   -1.3350   & -- &  --  \\
Gd$_{22}$Fe$_{68.2}$Co$_{9.8}$      &   1.0135   &   -1.2244   & -- &  --
\\
Gd$_{22}$Fe$_{68.2}$Co$_{9.8}$      &   0.9846     &     -0.7737       & -- &  --
\\
Gd$_{22}$Fe$_{74.6}$Co$_{3.4}$      &     0.9846     &   -0.8463   & -- &  --   \\
\hline
\underline{Tb$_{30}$Fe$_{70}$} & -- &  --  &    2.1506  &    -1.8385 \\
Tb$_{30}$Fe$_{70}$   &-- & --   &       1.7594    &     -1.4473  \\
Tb$_{30}$Fe$_{70}$   &-- & --   &       1.4952        &      -1.1831
\\
Tb$_{30}$Fe$_{70}$   &-- & --   &          1.4698           &       -1.1577       \\
\underline{Tb$_{29}$Fe$_{71}$}   &-- & --   &      2.0789      &   -1.8648    \\
Tb$_{27}$Fe$_{73}$  &-- & --  &       1.5835    &     -1.5093      \\
Tb$_{27}$Fe$_{73}$  &-- & --   &        1.1867    &     -1.1125     \\
Tb$_{24}$Fe$_{76}$  &-- &--   &       1.2641    &      -1.3452
\\
Tb$_{24}$Fe$_{76}$  &-- &--   &         1.1758         &      -1.2569         \\
Tb$_{22}$Fe$_{78}$   &-- & --   &     1.2789      &    -1.5007    \\
Tb$_{22}$Fe$_{78}$   &-- & --   &       1.1587           &    -1.3806
\\
Tb$_{22}$Fe$_{78}$   &-- & --   &          1.0965               &    -1.3183
\\
Tb$_{22}$Fe$_{78}$   &-- & --   &          0.9669 &        -1.1887
\\
\hline
\end{tabular}
\label{tab0}
\end{table}

\begin{figure}
\centerline{\psfig{file=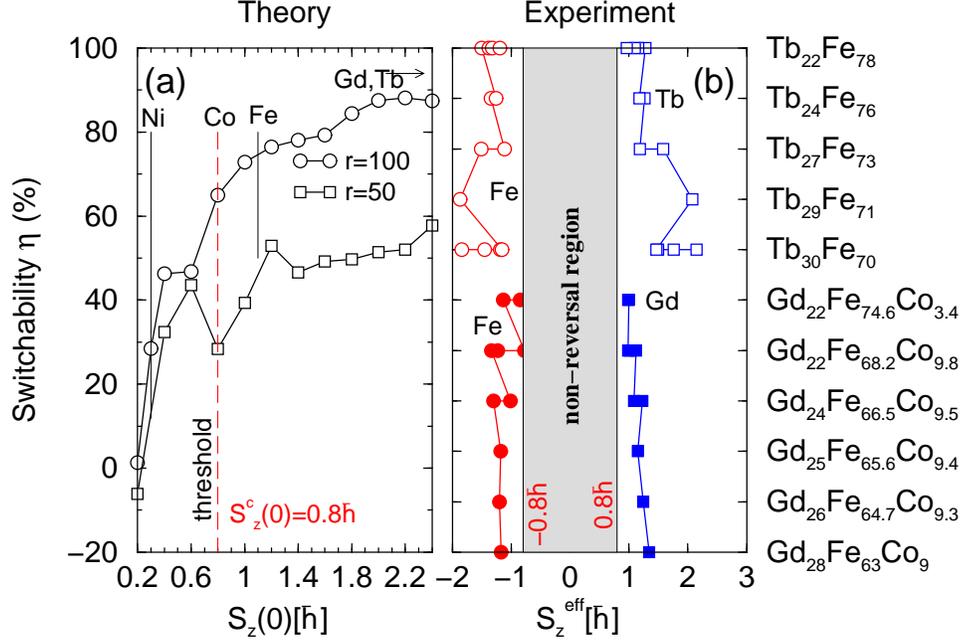,width=1\columnwidth,angle=0}}
\caption{ (a) Spin switchability versus the initial spin angular
  momentum $S_z(0)$ at the respective optimal laser field
  amplitudes.\protect\cite{epl16} The empty circles and boxes refer to
  the results with $r=100$ and $r=50$, respectively. The long-dashed
  line denotes the critical spin $S_z^c$.  Two thin vertical lines
  represent the spins for Ni and Fe. Co is on the border line, while
  Gd and Tb are way above $S_z^c$. The arrow on the top right refers
  to the fact that Gd and Tb have a much higher spin angular momentum.
  (b) Computed experimental effective spin angular momentum for each
  element in 11 GdFeCo and TbFe
  alloys.\protect\cite{hassdenteufel2015} Without exception, all
  elements have spin larger than $S_z^c$. Used with permission from
  EPL.}
\label{epl16fig3}
\end{figure}

\subsection{Dynamical simulation}


Different from the magnetic field-driven spin reversal, AOS relies on
a laser field to flip spin from one direction to another. However, to
describe such a process has been a big challenge. Most simulations
have been phenomenological,\cite{ostler2012} where laser fields are
treated as an effective magnetic field.\cite{mplb16} Ostler
\ete\cite{ostler2012} and Mentink \ete\cite{mentink2012} showed that
in GdFeCo, HID-AOS and HD-AOS depend on the laser intensity (electric
field squared), not the field helicity. As shown in our recent
study,\cite{prb17a} caution must be taken if the system has two spin
sublattices. Because the laser field is only active for one
sublattice, two sublattices are present and they separately allow the
laser field of either helicity to switch spin, so the final results
appear to the reader that AOS only depends on the laser intensity. If
one only has one spin orientation such as in CoPt ultrathin films, the
impact of the laser field, not just the laser intensity, appears.

In our view, the key to AOS theory is to include the microscopic
interaction between laser pulses and systems.  In 2016, we introduced
a new formalism that couples laser excitation to spin through
spin-orbit coupling, while the spin-spin interaction is described by
the Heisenberg exchange Hamiltonian, and electrons move inside a
harmonic potential.\cite{epl15} This model, though simple, overcomes
traditional difficulties that laser excitation and spin dynamics are
treated as two separate entities.  Specifically, the Hamiltonian is
\cite{epl15,epl16} \be H=\sum_i \left [\frac{{\bf p}_i^2}{2m}+V({\bf
    r}_i) +\lambda {\bf L}_i\cdot {\bf S}_i -e {\bf E}({\bf r}, t)
  \cdot {\bf r}_i\right ]-\sum_{ij}J_{ex}{\bf S}_i\cdot {\bf
  S}_{j}. \label{ham} \ee The summation is over all the lattice sites.
Here, the first term is the kinetic energy operator of the electron;
the second term is the harmonic potential energy operator with system
frequency $\Omega$; $\lambda$ is the spin-orbit coupling in units of
eV/$\hbar^2$; $ {\bf L}$ and $ {\bf S} $ are the orbital and spin
angular momenta in units of $\hbar$, respectively; and {\bf p} and
{\bf r} are the momentum and position operators of the electron,
respectively. Note that ${\bf L}$ is computed from ${\bf L}={\bf
  r}\times {\bf p}$, and there is no need to set up a different
equation for it.  The last term is the exchange interaction, and
$J_{ex}$ is the exchange integral in units of eV/$\hbar^2$.  Such a
Hamiltonian contains the necessary ingredients for AOS.  To our
knowledge, this is the only model that can generate spin reversal
without introducing a semiempirical effective magnetic field.

To demonstrate the power of this model, Figure \ref{epl16fig3}(a)
shows the theoretical result for the switchability $\eta$ as a
function of spin angular momentum $S_z$ in a ferromagnetic slab for
two radii of the laser radiation: one is the uniform radiation over
the entire sample, and the other is half the uniform
radius.\cite{epl16} The spin switchability is defined as $
\eta=\frac{S_z^f}{S_z(0)}\times 100\%,$ where $S_z^f$ is the final
spin angular momentum. It is clear that $\eta$ increases with
$S_z$. In order to realize AOS, there is a threshold value for the
spin momentum, $S_z^c=0.8\pm 0.2 \hbar$, that the material spin
momentum has to exceed. For instance, pure Ni can not exhibit AOS.  Co
is on the threshold.  As we noted before,\cite{epl16} Co-Pt granular
samples\cite{figueroa2014} have an effective spin magnetic moment per
$3d$ hole of 0.77 $\mu_{\rm B}$; since there are 2.49-2.62 holes, the
spin angular momentum is 0.96$\hbar$, satisfying this criterion.

What is even more interesting is that when we reduce the laser spot
size, the switch becomes more difficult. It is truly gratifying that
we predicted this result before the latest experimental results were
reported.\cite{vomir2017}


\begin{figure}
\centerline{\psfig{file=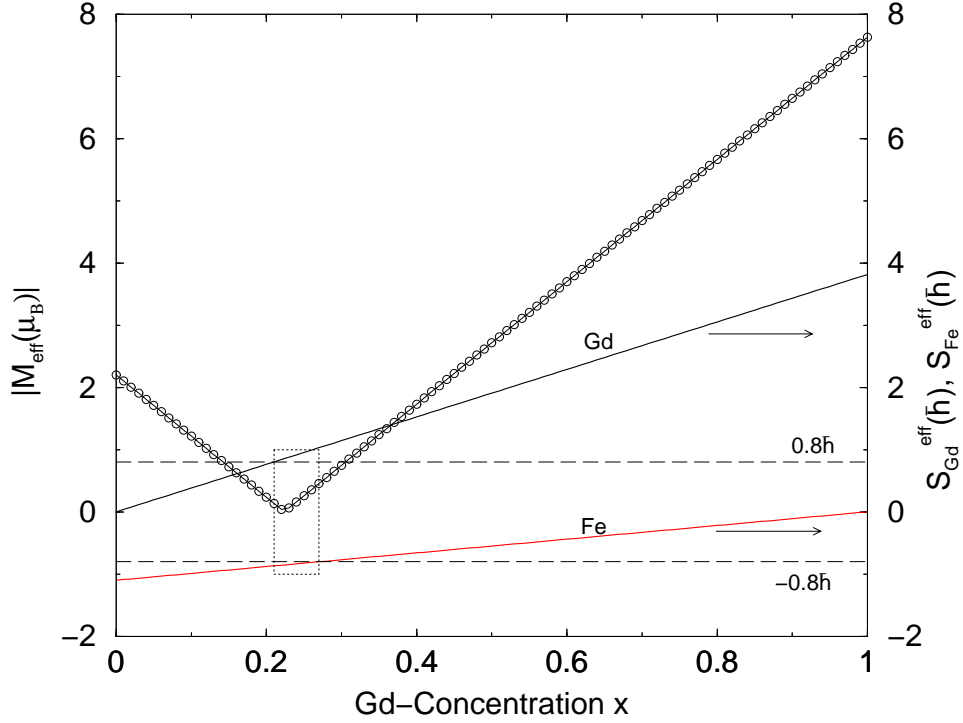,width=0.8\columnwidth,angle=270}}
\caption{ Effective spin moment change as a function of the Gd
  concentration $x$ (circles). The two solid lines represent the effective
  spin angular momenta for Gd and Fe (using the right axis). The two
  horizontal dashed lines denote the predicted critical spin angular
  momenta ($\pm 0.8\hbar$). The dotted line box highlights the narrow
  region of the Gd concentration where spin angular momentum satisfies
  our criterion and the spin reversal occurs.  }
\label{supfig0}
\end{figure}

We want to make connection with the above found spin angular momentum.
Figure \ref{epl16fig3}(b) shows a comparison with all 11 rare-earth
transition-metal ferrimagnets. These ferrimagnets show AOS. It is
remarkable that all the elements have spin angular momenta exceeding
the threshold value of $0.8\hbar$. To understand the window of
concentration $x$, Figure \ref{supfig0} plots the spin moment as a
function of the Gd concentration $x$. Two horizontal dashed lines set
the spin threshold for each element. The rectangular box delineates
the window for all-optical switching. The agreement between experiment
and theory gives confidence that the model works reasonably well.

\section{Simple Theory for All-Optical Switching}

Despite a decade of investigation, our understanding is still very
limited. As discussed above, the majority of theoretical research has
been phenomenological.\cite{mplb16} This calls for a systematic
experimental investigation by tuning both system- and laser-specific
parameters.  Before one can pin down the origin of AOS, it is
necessary to develop a many-to-one correspondence between the proposed
mechanisms (see Table \ref{tab1}) and more fundamental
interactions. However, the majority of the proposed mechanisms are
very difficult to attribute to a single interaction.  We speculate
that the requirement of compensation points can be mapped to the
requirement of a sizable spin moment, while the domain size criterion
could be a result associated with radiation size as shown in our
recent study\cite{jpcm16} and the Vomir \et
observation.\cite{vomir2017}

By contrast, theoretically the inverse Faraday effect is more mature
and has an existing theory developed for a nonabsorbing
medium.\cite{pershan1966} According to Shen,\cite{ren1984} the optical
field-induced magnetization is \be \Delta M_{F}=i \frac{\partial
  \chi_{xy}}{\partial H_0}(|E_+|^2-|E_-|^2), \ee where $H_0$ is the
static magnetic field and $\chi_{xy}$ is the
susceptibility. Interestingly, this expression is different from the
commonly used one with the field product ${\bf E}\times {\bf E}$.
Note that Shen considered a nonabsorbing medium. Here $E_+$ and $E_-$
are $(E_x-iE_y)/\sqrt{2}$ and $(E_x+iE_y)/\sqrt{2}$,
respectively.\cite{pershan1966} For linearly polarized light,
$E_+=E_-$, so $\Delta M_{F}=0$; for circularly polarized light,
$\Delta M_{F}$ is nonzero and changes signs from the left-circularly
polarized light to the right one.  $\chi_{xy}$'s frequency dependence
is not considered.\cite{pershan1966} To apply the above equation to an
absorbing material, we use our susceptibility \cite{epl15,mplb16} for
a single site. Our model\cite{epl15} does not have an external field,
but has a spin-orbit coupling (see Eq. (\ref{ham})), so the spin
angular momentum plays the same role as the magnetic field. Note that
in the SI unit system, the linear susceptibility $\chi^{(1)}_{xy}$ has
no unit.  We take the derivative of $\chi^{(1)}_{xy}$ with respect to
$S_z$.  $\chi^{(1)}_{xy}$ has $S_z$ in its numerator and denominator,
so the derivative is complicated. For simplicity, we only include the
derivative from the numerator. We have the magnetization change \be
\Delta M_{F} \propto \frac{Ne^2}{\epsilon_0 m}\frac{2\lambda
  \omega}{(\Omega^2-\omega^2-\lambda^2S_z^2)^2-(2\lambda S_z\omega)^2}
(|E_+|^2-|E_-|^2), \label{m} \ee where $\lambda$ is the spin-orbit
coupling and $S_z$ is the spin angular momentum. Other variables can
be found in our prior paper.\cite{epl15} This relation highlights the
importance of the spin-orbit coupling and reveals how circularly
polarized light may affect spins.

\begin{figure}
\centerline{\psfig{file=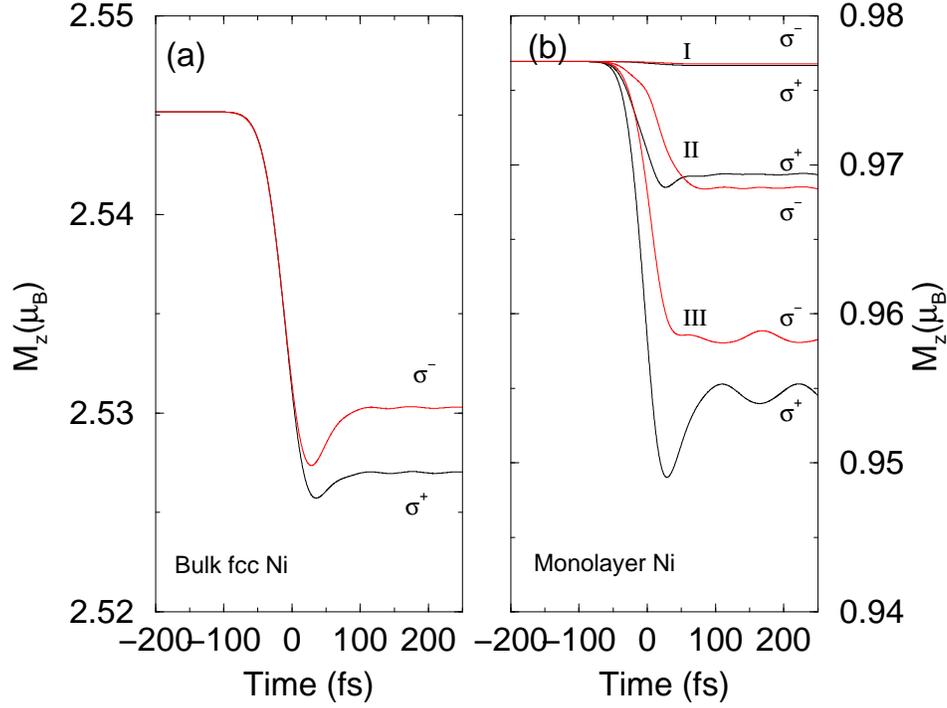,width=1\columnwidth,angle=0}}
\caption{ (a) Spin moment change as a function of time in fcc Ni for
  the left ($\sigma^-$) and right ($\sigma^+$) circularly polarized
  light.  The laser pulse duration is $\tau=60$ fs, the photon energy
  is $\hbar\omega=2$ eV, and the field amplitude is $A_0=0.0099 \rm
  Vfs/\AA$. (b) Spin moment change as a function of time in monolayer
  Ni for  left ($\sigma^-$) and right ($\sigma^+$) circularly
  polarized light. We choose three sets of laser parameters. 
I:  $\tau=48$ fs,   $\hbar\omega=1.6$ eV  and $A_0=0.0030 \rm  Vfs/\AA$; 
II:  $\tau=48$ fs,   $\hbar\omega=1.6$ eV and $A_0=0.030 \rm  Vfs/\AA$; 
III:  $\tau=48$ fs,   $\hbar\omega=1.55$ eV and $A_0=0.030 \rm  Vfs/\AA$.
}
\label{ni}
\end{figure}

The above results are reproduced in our first-principles calculation.
We carry out a lengthy calculation using the time-dependent Liouville
density functional theory\cite{jpcm16} for both bulk fcc Ni and
monolayer Ni. We employ both $\sigma^+$ and $\sigma^-$.  We choose a
60-fs laser pulse with photon energy 2 eV and vector potential
amplitude $0.009\rm Vfs/\AA$.  Figure \ref{ni}(a) shows that the laser
helicity does affect the spin moment change.\cite{lu2018}  We see that $\sigma^+$
can demagnetize Ni more than $\sigma^-$. In the simulation, we adopt a
simple cubic structure for fcc Ni, so there are four Ni atoms in our
cell. Figure \ref{ni}(b) shows the results for a Ni(001) monolayer
for three sets of laser parameters, labeled by I, II and III. For I,
we use $\tau=48$ fs, $\hbar\omega=1.6$ eV and $A_0=0.0030 \rm
Vfs/\AA$. We see that under such a weak laser pulse the
demagnetization is very small. However, the difference in spin moment
between $\sigma^+$ and $\sigma^-$ excitation is also visible. Set II
shows an increase in the laser field amplitude to $\tau=48$ fs,
$\hbar\omega=1.6$ eV and $A_0=0.030 \rm Vfs/\AA$.  We find that
different helicity induces different demagnetization. Even the
demagnetization time is different, 27 fs for $\sigma^+$ and 84 fs for
$\sigma^-$. It is not always true that one kind of helicity dominates
all the time. We also investigate how the photon energy affects the
demagnetization. We decrease the photon energy to set III with the
laser parameters $\tau=48$ fs, $\hbar\omega=1.55$ eV and $A_0=0.030
\rm Vfs/\AA$. Figure \ref{ni}(b) shows that in comparison with set II,
the monolayer demagnetizes more for the same laser helicity as we
decrease the photon energy. In addition, there is a strong oscillation
in the spin moment.  In this case, $\sigma^+$ induces a more
pronounced change.

While the above analytic and numerical results are insightful in
themselves, they miss some crucial experimental and numerical
findings.  For instance, numerically we find that linearly polarized
light can switch spins as well if the laser field amplitude becomes
stronger,\cite{epl15,epl16} but Eq. (\ref{m}) gives zero for linearly
polarized light. This highlights that this analytic expression, which
is obtained under perturbation theory, does not catch the actual spin
reversal completely. Our first-principles result only shows the
demagnetization, not switching. The only way that we can get true spin
reversal is to use the Hamiltonian of Eq. (\ref{ham}). We find that
the actual switching may result from the spin-orbit
torque,\cite{mplb16} \be {\bf \tau}= \lambda {\bf L}\times {\bf S},
\ee where ${\bf L}$ is the orbital angular momentum.  This was first
tested by Ren \et experimentally.\cite{ren2017}

\begin{figure}
\centerline{\psfig{file=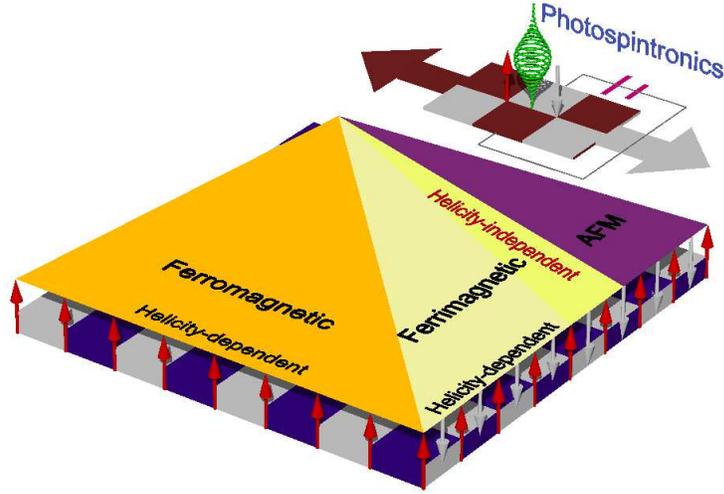,width=0.8\columnwidth,angle=0}}
\caption{ Phase diagram of AOS. Switchings in ferromagnetic (orange
  triangle) and weak ferrimagnetic materials (light yellow triangle)
  are always helicity-dependent. Helicity-independent switching
  (yellow triangle) occurs in a narrow region when the sublattice
  spins approach the antiferromagnetic limit.\protect\cite{prb17a} }
\label{phase}
\end{figure}

Recently we constructed a phase diagram, which is reproduced in
Fig. \ref{phase}.\cite{prb17a} We suggest that all the AOS materials
should be classified into three categories: ferromagnetic, weak and
strong ferrimagnetic.  In both ferromagnets and weak ferrimagnets,
only one spin orientation is present or dominant, and switching is
helicity-dependent.  For strong ferrimagnets, since both sublattices
have a strong magnetic moment, circularly polarized light with
different helicities can switch spins. This is potentially useful for
future device design.


\section{Importance of Perpendicular Magnetic Anisotropy: Emergence of
  Orbital Angular Momentum}

From the above discussion, it should be clear that one needs a large
spin moment for each constituent in a compound in order to switch
spins optically. However, if this is the sole criterion, then the
strongest man-made magnet Nd$_{2}$Fe$_{14}$B would be the best
candidate for AOS. However, this does not happen.  One crucial element
among all the six common themes is perpendicular magnetic anisotropy
(PMA).

We decide to first examine the universal presence of perpendicular
magnetic anisotropy (PMA) in nearly all the AOS materials. From our
prior investigation,\cite{jpcm17b} we find that PMA has an
unparalleled advantage over other spin configurations. The presence of
PMA demands a nonzero orbital angular momentum. It is well known that
in thin films, the surface contribution, which favors PMA, becomes
larger.  However, in solids the orbital angular momentum is largely
quenched by the crystal field. This statement is based on the symmetry
argument. If two spherical harmonics $Y_{lm}$ and $Y_{l\bar{m}}$ in a
Bloch state have the same weight, then the summation of the orbital
angular momentum expectation value $\langle
Y_{lm}|l_z|Y_{lm}\rangle$+$\langle
Y_{l\bar{m}}|l_z|Y_{l\bar{m}}\rangle$ is zero, where $l_z$ is the
orbital angular momentum operator along the $z$ axis. With the
presence of spin-orbit coupling, the equivalency of the two harmonics
breaks down. The level of breakdown depends on the strength of the
spin-orbit coupling. If we go back to Table \ref{tab1}, we notice that
rare-earth and $5d$ elements are present in all the materials. They
are known for their strong spin-orbit coupling. This implies a sizable
orbital momentum present in those compounds.

However, one often argues that since Gd atom has a half-filled $4f$
shell, its orbital angular momentum is zero.  However, this is no
longer the case in solids. Even in pure bulk Gd, the orbital momentum
is nonzero. We found\cite{prb17a} that hcp Gd has (0.00002, -0.00006,
0.12166)$\mu_B$, which is larger than that in Ni. Only if we include
the Hubbard $U$ term, can we quench it to nearly zero (-0.00006
0.00007, 0.01716)$\mu_B$, where we use $U=0.4926$Ry and $J=0.051$Ry.
These $+U$ treatments rectify correlation effects in the ground state
and do not change spin moments (compare 7.55298$\mu_B$ with $+U$ with
7.44249$\mu_B$ without $+U$), but worsen other properties, in
particular excitation processes, as shown recently.\cite{prb17a} The
orbital angular momentum is one of those properties for which the
GGA+U treatment does not work.  Jang \et \cite{jang2016} recently
demonstrated the existence of orbital ordering in GdB$_4$. Gd loses
three $5d$ electrons to B, so it becomes Gd$^{3+}$ with its $4f$
orbital half-filled.  They found the orbital order is strongly coupled
with the antiferromagnetic spin order. This matches the scenario in
GdFeCo, where antiferromagnetic ordering is also present. Therefore,
there is no contradiction regarding the importance of orbital angular
momentum and the presence of spin-orbit coupling.

\begin{figure}
\centerline{\psfig{file=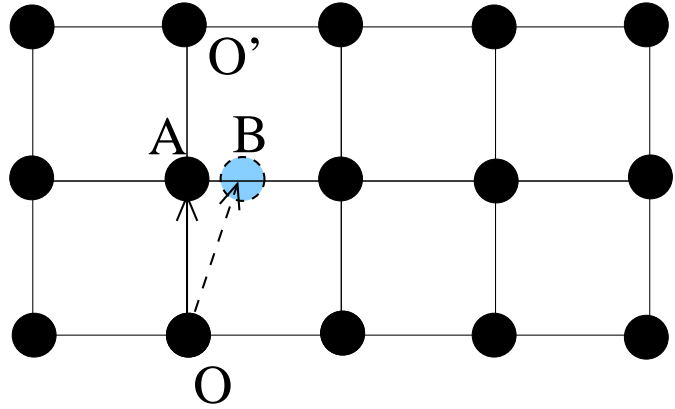,width=0.7\columnwidth,angle=0}}
\caption{Orbital angular momentum of lattice vibration in solids is
  difficult to define. The electron interacts with the lattice mainly
  through the energy exchange, instead of momentum exchange. 
  }
\label{orb}
\end{figure}

Different from the ground-state property, the orbital angular momentum
can be enhanced during laser excitation, during which the laser
helicity information is encoded. In other words, the electron orbital
angular momentum stores the helicity information. This information can
not be easily erased because the coupling between electron and lattice
subsystems is through energy exchange. The lattice orbital angular
momentum change, i.e., the lattice displacement $\Delta {\bf R}_{\rm
  lattice}$ $\times$ the lattice momentum ${\bf P}_{\rm lattice}$, is
tiny, because $\Delta {\bf R}_{\rm lattice}$ and ${\bf P}_{\rm
  lattice}$ are mostly along the same direction. In fact, in a
one-dimensional system, the orbital angular momentum of the lattice is
always zero, since the position and momentum of the lattice are always
in the same or in opposite directions.  Caution must be taken for 2-
and 3-dimensional systems when one tries to define the angular
momentum for a lattice. The orbital angular momentum of an atom
depends on the reference point and, if not treated properly, it
becomes ambiguous.

In Fig. \ref{orb}, we show an example, where we treat the motion of
atoms classically.  Suppose that the equilibrium position of an atom
in a square lattice is at $A$. It linearly moves to $B$ with a certain
velocity. Such a linear motion should have zero angular
momentum. However, if we choose a reference point at $O$, then the
angular momentum is nonzero. If we choose $O'$ as the reference point,
then the angular momentum changes sign.  Besides, in general, atoms in
a lattice experience a crystal potential that is translationally
invariant, and do not orbit. This is different from molecular
crystals, such as C$_{60}$ solids, where C$_{60}$ spins rapidly at
room temperature.  For this reason, the traditional solid state theory
does not invoke the orbital angular momentum transfer between
electrons and lattices. The electron-phonon interaction relies on the
energy transfer, where the phonon frequency is renormalized.

\section{Path to Single-Shot Switching in Ferromagnets}

\begin{figure}
\centerline{\psfig{file=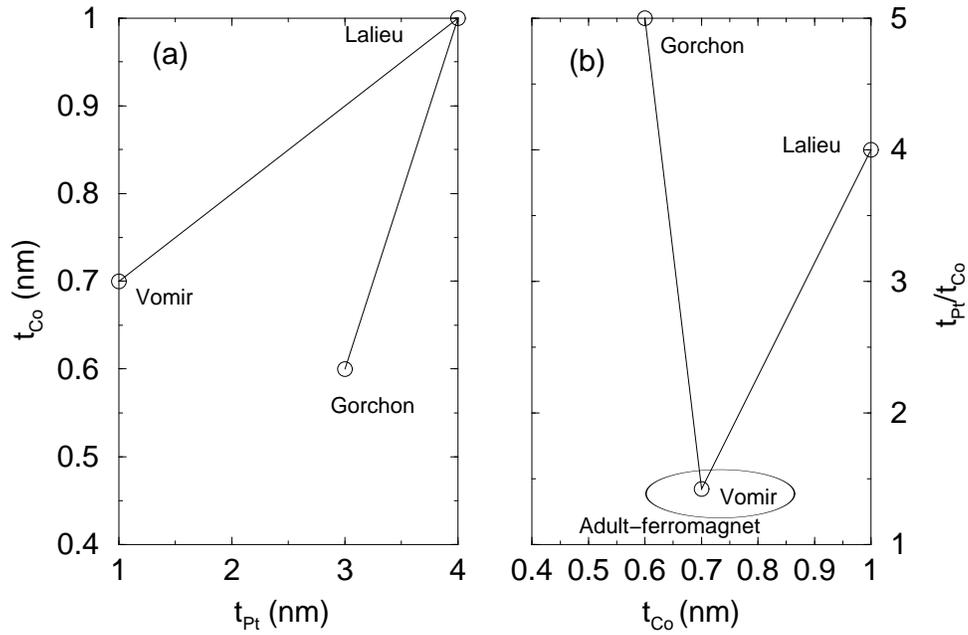,width=0.7\columnwidth,angle=270}}
\caption{Importance of layer thickness on  single-shot spin
  switching.  (a) Co layer thickness $t_{\rm Co}$ versus Pt thickness
  $t_{\rm Pt}$ for three experiments. (b) Ratio $t_{\rm Pt}$/$t_{\rm
    Co}$ as a function of $t_{\rm Co}$.  }
\label{copt}
\end{figure}

Up to now, single-shot spin switching is dominated by a single
material, ferrimagnet GdFeCo, which can be driven optically or
electrically.\cite{yang2017} Many newly discovered
materials\cite{mangin2014} do not have this property. This will
certainly limit future applications, as the majority of materials used
for magnetic storage are ferromagnetic.  Fortunately, recent studies
show there are at least three experiments
\cite{lalieu2017,gorchon2017,vomir2017} that allow single-shot
switching in ferromagnets. Two of them still rely on Gd and GdFeCo,
which will be called rare-earth-assisted ferromagnets, or RE-parenting
ferromagnets. Only one is independent, an adult-ferromagnet. All of
the samples are Co/Pt ultrathin layers. While it is really premature
to make sense out of a limited set of data, we wonder what makes Co/Pt
so unique in this aspect. It is also interesting to note that the same
material is used for spin transport.\cite{baek2018}

These experiments do not provide the detailed structure
characterization and magnetic properties, except that all of them are
found to have PMA.  Figure \ref{copt}(a) plots the thickness of a Co
layer versus that of a Pt layer. We notice again that these films are
extremely thin, so the surface contribution dominates. We caution
again that within such a thickness, the nominal thickness may differ
from the actual one. There is no clear trend among those three
experiments. But it is known that the reduced dimensionality and
imperfections affect PMA greatly.\cite{brahimi2016} The interfacial
effects affect magnetic relaxations in Co/Pt
multilayers.\cite{yuan2003} In addition, very large domain wall
velocities were found.\cite{pham2016} Atoms across the interface may
segregate into different regions in nanoparticles.\cite{liu2016} The
presence of PMA further indicates the presence of orbital angular
momentum.\cite{liu2017} It also matters that the arrangement of atoms
across the interface affects the magnetic properties.  One can see
this from the experiments too.  Figure \ref{copt}(b) plots the
thickness ratio versus the Co layer thickness. The data of Vomir
\ete\cite{vomir2017} has the smallest ratio. We wonder whether this
makes their sample as the only one sample for a single-shot AOS.  In
addition, the fact that Vomir \ete\cite{vomir2017} obtained AOS at a
much longer time indicates the damping of magnetization must be very
small.  This matches our expectation. Since the spin-orbit torque and
spin damping result from the same spin-orbit coupling, a reduced
spin-orbit coupling must have a longer switching time. Recent
theoretical studies at Co/Pt interfaces reveal additional information
about the Dzyaloshinskii-Moriya interaction at disordered
interfaces.\cite{zimmermann2018}

\section{Conclusions}

It is an exciting time to investigate all-optical spin
switching. Progress has been very impressive. It is generally agreed
that the microscopic picture of AOS is complex.  AOS becomes strongly
material-dependent, but what material's properties the actual
dependence depends on is unclear.  There are many cases where opposite
situations appear simultaneously.  Therefore, how to unify different
fractions of underlying mechanisms has dominated much of the latest
research. Yet, the theory is mostly phenomenological and far from
perfect. The irony is that if one carries out a first-principles
calculation for a ferromagnet, the effective switching field is tiny,
and the laser simply can not switch spins. A stronger laser only
demagnetizes a sample, but does not switch the spin direction.  The
origin of the theoretical failure is also unknown. There is a great
need to construct a minimal model, without introducing an effective
magnetic field, to simulate AOS. This could be helpful to understand
how magnetic domains affect the switching. What we miss is the
structure characterization of samples. We believe with additional
experimental and theoretical investigations that AOS can be put into
practical usage.\cite{chen2017,davies2018}


\section*{Acknowledgments}

We greatly appreciate Dr. Jeff Bokor (Berkeley, US) for reviewing our
paper with helpful comments and remarks on the computer clock
frequency. We greatly benefited from the first two references that he
provided in the first paragraph of this paper.  We would like to thank
Dr. Theo Rasing for reviewing our draft. Both Drs. Bokor and Rasing
noted that AOS could occur in samples with in-plane magnetic
anisotropy.  We are extremely grateful to Dr. Manfred Albrecht
(Ausburg, Germany) who has thoroughly examined our paper and provided
helpful suggestions. Some of the changes are directly adopted from his
comments. To support the open code initiative, we attach our computing
code. The spin switching code will be presented in a book that we are
completing.

 This work was supported by the U.S. Department of Energy under
 Contract No. DE-FG02-06ER46304. M. S. Si was supported by the NSFC of
 China under No. 11874189.  Part of the work was done on Indiana State
 University's quantum cluster and high-performance computers.  The
 research used resources of the National Energy Research Scientific
 Computing Center, which is supported by the Office of Science of the
 U.S. Department of Energy under Contract No. DE-AC02-05CH11231. We
 would like to thank Anthony Froehlich for assistance with the
 figures.

\appendix

\section*{Appendix  A. Computer codes for 
 spin angular momentum in amorphous materials}

\begin{verbatim}
! The following code was written by Guoping Zhang, January 2016 at
! Indiana State University, Terre Haute, IN 47809, U.S.A.

! The code development was supported by U. S. Department of Energy under
! Contract No. DE-FG02-06ER46304.

! This code allows one to compute spin angular momenta for amorphous
! materials.

! Anyone can freely share this code. But please keep the headings. 
! Copyrighted by Guoping Zhang. 

      implicit double precision(a-z)     
      character *15 sample(2)
      character *2 element(2)
      double precision percent(2,2),magnet(2)
      integer i
!     spin: the spin for each element
!     spin_eff: the effective spin which is computed by the spin x percent
      double precision spin(2),spin_eff(2,2),spin_hbar(2,2)
      double precision vol_(2)

!     this is  Terbium lattice constant. I only use Terbium bulk for all
!     alloys.
!      ml=3.03d0 ! M_l for Terbium.
      a=360.1d0*1d-2
      b=a
      c=569.36d-2
      vol_tb=a**2*c*dsqrt(3d0)/2d0
      vol_fe=2.8665d0**3!/2d0
      vol_co=(2.5071d0)**2*4.0695d0*dsqrt(3d0)/2d0
      vol_gd=(3.636d0)**2*5.7826*dsqrt(3d0)/2d0
      write(*,*)vol_tb,vol_fe,vol_co,vol_gd
      open(2,file='output.data')
!      write(2,*)'\hline'
      open(1,file='input.data')
      read(1,*)

 111  read(1,*,end=110)sample(1),percent(1,1),percent(2,1),element(1)
     $     ,element(2),magnet(1)

      read(1,*)sample(2),percent(1,2),percent(2,2),element(1),element(2)
     $     ,magnet(2)

      do i=1,2
         vol_(i)=0d0
         write(*,*)i,vol_(i)
         write(*,*)i,element(i)
         if(element(i).eq.'gd')vol_(i)=vol_gd
         if(element(i).eq.'tb')vol_(i)=vol_tb
         if(element(i).eq.'fe')vol_(i)=vol_fe
         if(element(i).eq.'co')vol_(i)=vol_co
      enddo

      do i=1,2
         write(*,*)i,vol_(i)
      enddo
!     stop

      call treat(sample,magnet,vol_,percent,spin,spin_eff,spin_hbar)

      write(*,*)sample(1),percent(1,1),percent(2,1),element(1)
     $     ,'   ',element(2),magnet(1)

      write(*,*)sample(2),percent(1,2),percent(2,2),element(1)
     $     ,'   ',element(2),magnet(2)
      write(*,302)element(1),spin(1),'[uB]'
      write(*,302)element(2),spin(2),'[uB]'

      write(2,3021)element(1),'&',spin(1),'$\mu_{B}$','&','--&','\\'
      write(2,3021)element(2),'&',spin(2),'$\mu_{B}$','&','--&','\\'

 3021 format(1x,2a,f10.4,4a)

 302  format(1x,a,f10.4,a)
      do i=1,2
         write(*,*)sample(i),spin_eff(1,i),spin_eff(2,i),'[uB]'
         write(*,*)sample(i),spin_hbar(1,i),spin_hbar(2,i),'[hbar]'
 3022    format(1x,3a,f10.4,a,f10.4,a)
         write(2,3022)sample(i),'&','&',spin_hbar(1,i),'$\hbar$&'
     $        ,spin_hbar(2,i),'$\hbar$ \\ '
      enddo
      write(2,*)'\hline'
      goto 111
 110  close(1)
      close(2)

      stop
      end
      subroutine treat(sampl,magnet,vol,percent,spin,spin_eff,spin_hbar)
!      implicit double precision (a-h,o-z)
      implicit none
      integer i,j
!     INPUT
      character *15 sampl(2)
!     percent: concentration for compound 1 percent(*,1)
      double precision vol(2),percent(2,2),magnet(2)
      double precision to_uB
!     output
!     spin: the spin for each element
!     spin_eff: the effective spin which is computed by the spin x percent
      double precision spin(2),spin_eff(2,2),spin_hbar(2,2)
      
!     ax+by=s for compound (1). 
!     s: the magnetization for compound (1)
!     a: percentage of element 1 = x1 in Eq. (3). For Gd_xFe_yCo, a=x
!     b: percentage of element 2 = (1-x1) in Eq. (3). For Gd_xFe_yCo, b=y
!     x: M_R in eq. (3) 
!     y: M_T in eq. (3) 

!     cx+dy=t for compound (2)
!     t: the magnetization for compound (2)
!     c,d have the same meaning as a and b, except for compound (2)
!     x,y: see above
      double precision a,b,c,d,s,t,x,y,volume
      
      to_uB=1d-3/9.274d0        !this converts kA/m to u_B
!     ax+by=s; cx+dy=t

!     this is for compound 1
      a=percent(1,1)            !element 1 
      b=percent(2,1)            !element 2
!     volume is effective cell volume in unit of Angstrom cubic 
      volume=a*vol(1)+b*vol(2)
!     convert magnetization to magnetic moment, then convert to u_B
      s=magnet(1)*volume*to_uB

!     this is for compound 2
      c=percent(1,2) 
      d=percent(2,2)
      volume=c*vol(1)+d*vol(2)
!     convert magnetization to magnetic moment, then convert to u_B
      t=magnet(2)*volume*to_uB

!     the following is the element's spin moment in unit of u_B
      y=(a*t-c*s)/(a*d-c*b)
      x=(s-b*y)/a
      
      spin(1)=x
      spin(2)=y

!     effective spin for compound 1
      spin_eff(1,1)=spin(1)*a   !element 1
      spin_eff(2,1)=spin(2)*b   !element 2

!     effective spin for compound 2
      spin_eff(1,2)=spin(1)*c   !element 1
      spin_eff(2,2)=spin(2)*d   !element 2

!     convert from uB to hbar, for zero orbital angular momentum, the
!     conversion factor is 2, but if the element has a nonzero orbital
!     angularl moment. The conversion needs some caution. 
      
      do i=1,2
         do j=1,2
            spin_hbar(i,j)=spin_eff(i,j)/2d0
         enddo
      enddo
      end     
\end{verbatim}

\section*{Appendix B. Sample input files}

This file is called {\sf  input.data}. The experimental remanence is
from the paper by Hassdenteufel \ete\cite{hassdenteufel2015}

\begin{verbatim}
#compound   percentages   elements  remanent (kA/m)
Tb30Fe70------- 0.30 0.70  tb fe     162.3d0
Tb29Fe71------- 0.29 0.71  tb fe     112.64d0 
\end{verbatim}

\section*{Appendix C. Sample output file}

This file is called {\sf  output.data}.
\begin{verbatim}
 tb&   14.3375$\mu_{B}$&--&\\
 fe&   -5.2529$\mu_{B}$&--&\\
 Tb30Fe70-------&&    2.1506$\hbar$&   -1.8385$\hbar$ \\ 
 Tb29Fe71-------&&    2.0789$\hbar$&   -1.8648$\hbar$ \\ 
 \hline
\end{verbatim}
The above results are highlighted  in Table \ref{tab0}. See the first line
starting with \tbfe{30}{70} and the line starting with \tbfe{29}{71}.


\begin{thebibliography}{0}

\bibitem{theis2010a}T. N. Theis and P. M. Solomon,
{It's time to reinvent the transistor!} Science {\bf 327}, 1600
(2010).

\bibitem{theis2010b}T. N. Theis and P. M. Solomon,
{In quest of the ``next switch'': Prospects for greatly reduced power
dissipation in a successor to the silicon field-effect
transistor}, Proc. IEEE {\bf 98}, 2005 (2010).

\bibitem{stanciu2007}C. D. Stanciu, F. Hansteen, A. V. Kimel, A. Kirilyuk,
A. Tsukamoto, A. Itoh, and Th. Rasing, All-optical magnetic
recording with circularly polarized light, \prl {\bf 99}, 047601
(2007).

\bibitem{eric} E. Beaurepaire, J. C. Merle, A. Daunois, and J.-Y. Bigot,
{Ultrafast spin dynamics in ferromagnetic nickel},
Phys.  Rev. Lett. {\bf 76}, 4250 (1996).

\bibitem{hansen1988} P. Hansen and K. Witter, The role of the
compensation temperature in thermomagnetic switching, IEEE
Trans. Magn. {\bf 24}, 2317 (1988).

\bibitem{ostler2012}T. A. Ostler, J. Barker, R. F. L. Evans,
R. W. Chantrell, U. Atxitia, O. Chubykalo-Fesenko, S. El Moussaoui,
L. Le Guyader, E. Mengotti, L. J. Heyderman, F. Nolting,
A. Tsukamoto, A. Itoh, D. Afanasiev, B. A. Ivanov,
A. M. Kalashnikova, K. Vahaplar, J. Mentink, A. Kirilyuk,
Th. Rasing, and A. V. Kimel, {Ultrafast heating as a sufficient
stimulus for magnetization reversal in a ferrimagnet},
Nat. Commun. {\bf 3}, 666 (2012).

\bibitem{lalieu2017} M. L. M. Lalieu, M. J. G. Peeters,
S. R. R. Haenen, R. Lavrijsen, and B. Koopmans, Deterministic
all-optical switching of synthetic ferrimagnets using single
femtosecond laser pulses, Phys. Rev. B {\bf 96}, 220411 (2017).

\bibitem{gorchon2017}J. Gorchon, C.-H. Lambert, Y. Yang, A. Pattabi,
R. B. Wilson, S. Salahuddin, and J. Bokor, {Single shot
ultrafast all optical magnetization switching of ferromagnetic
Co/Pt multilayers}, Appl. Phys. Lett. {\bf 111}, 042401 (2017).

\bibitem{vomir2017}M. Vomir, M. Albrecht, and J.-Y. Bigot, Single shot
all optical switching of intrinsic micron size magnetic domains of a
Pt/Co/Pt ferromagnetic stack, Appl. Phys. Lett. {\bf 111}, 242404
(2017).

\bibitem{finazzi2013}M. Finazzi, M. Savoini, A. R. Khorsand,
A. Tsukamoto, A. Itoh, L. Duo, A. Kirilyuk, Th.Rasing, and M. Ezawa,
Laser-induced magnetic nanostructures with tunable topological
properties, Phys. Rev. Lett.  {\bf 110}, 177205 (2013).

\bibitem{rasingreview}A. Kirilyuk, A. V. Kimel, and Th. Rasing,
{Ultrafast optical manipulation of magnetic order},
Rev. Mod. Phys. {\bf 82}, 2731 (2010). Erratum Rev. Mod. Phys. {\bf
88}, 039904 (2016).

\bibitem{ourreview}G. P. Zhang, W. H\"ubner, E.  Beaurepaire, and
J.-Y. Bigot, Laser-induced ultrafast demagnetization: Femtomagnetism,
A new frontier? Topics Appl. Phys.  {\bf 83}, 245 (2002).

\bibitem{mplb16}G. P. Zhang, T. Latta, Z. Babyak, Y. H. Bai, and
T. F. George, All-optical spin switching: A new frontier in
femtomagnetism --  A short review and a simple theory, Mod. Phys.
Lett. B  {\bf 30}, 1630005 (2016).

\bibitem{elhadri2017}M. S. El Hadri, M. Hehn, G. Malinowski, and
S. Mangin, {Materials and devices for all-optical helicity-dependent
switching}, J. Phys. D: Appl. Phys. {\bf 50}, 133002 (2017).

\bibitem{lu2018}X. Lu, X. Zou, D. Hinzke, T. Liu, Y. Wang, T. Cheng,
J. Wu., T. A. Ostler, J. Cai, U. Nowak, R. W. Chantrell, Y. Zhai,
and Y. Xu, Roles of heating and helicity in ultrafast all-optical
magnetization switching in TbFeCo, Appl. Phys. Lett. {\bf 113},
032405 (2018)

\bibitem{liu2015}T.-M. Liu \ete, Nanoscale confinement of
all-optical magnetic switching in TbFeCo - Competition with
nanoscale heterogeneity, NanoLett. {\bf 15}, 6862 (2015).

\bibitem{kimel2004} A. V. Kimel, A. Kirilyuk, A. Tsvetkov,
R. V. Pisarevm, and Th. Rasing, {Laser-induced ultrafast spin
reorientation in the antiferromagnet TmFeO$_3$}, Nature {\bf 429},
850 (2004).

\bibitem{kimel2005}A. V. Kimel, A. Kirilyuk, P. A. Usachev,
R. V. Pisarev, A. M. Balbashov, and Th. Rasing, {Ultrafast
non-thermal control of magnetization by instantaneous
photomagnetic pulses}, Nature {\bf 435}, 655 (2005).

\bibitem{kimel2009}A. V. Kimel, B. A. Ivanov, R. V. Pisarev,
P. A. Usacgev, A. Kirilyuk, and Th. Rasing, {Inertiea-driven spin
switching in antiferromagnets},
Nat. Phys. {\bf 5}, 727 (2009).

\bibitem{jin2010}Z. Jin, H. Ma, L. Wang, G. Ma, F. Guo, and J. Chen,
{Ultrafast all-optical magnetic switching in NaTb(WO$_{4}$)$_2$},
App. Phys. Lett. {\bf 96}, 201108 (2010).

\bibitem{hohlfeld2009}J. Hohlfeld, C. D. Stanciu, and A. Rebel,
{Athermal all-optical femtosecond magnetization reversal in GdFeCo},
Appl. Phys. Lett. {\bf 94}, 152504 (2009).

\bibitem{vahaplar2009} K. Vahaplar, A. M. Kalashnikova, A. V. Kimel,
D. Hinzke, U. Nowak, R. Chantrell, A. Tsukamoto, A. Itoh,
A. Kirilyuk, and Th. Rasing, {Ultrafast path for optical
magnetization reversal via a strongly nonequilibrium state},
Phys. Rev. Lett. {\bf 103}, 117201 (2009).

\bibitem{steil2011}D. Steil, S. Alebrand, A. Hassdenteufel,
M. Cinchetti, and M. Aeschlimann, {All-optical magnetization
recording by tailoring optical excitaiton parameters},
Phys. Rev. B {\bf 84}, 224408 (2011).

\bibitem{ohkochi2012}T.  Ohkochi, H.  Fujiwara, M.  Kotsugi,
A. Tsukamoto, K. Arai, S. Isogami, A. Sekiyama, J. Yamaguchi,
K. Fukushima, R. Adam, C. M. Schneider, T. Nakamura, K.  Kodama,
M. Tsunoda, T. Kinoshita, and S.  Suga, {Microscopic and
spectroscopic studies of light-induced magnetization switching
GdFeCo facilitated by photoemission electron microscopy},
Jpn. J. Appl. Phys. {\bf 51}, 073001 (2012).

\bibitem{vahaplar2012} K. Vahaplar, A. M. Kalashnikova,
A. V. Kimel, S. Gerlach, D. Hinzke, U. Nowak, R. Chantrell,
A. Tsukamoto, A. Itoh, A. Kirilyuk, and Th. Rasing, {All-optical
magnetization reversal by circularly polarized laser pulshses:
Experiment and multiscale modeling}, Phys. Rev. B {\bf 85}, 104402
(2012).

\bibitem{khorsand2012}A. R. Khorsand, M. Savoini, A. Kirilyuk,
A. V. Kimel, A. Tsukamoto, A. Itoh, and Th. Rasing, {Role of
magnetic circular dichroism in all-optical magnetic recording},
Phys. Rev. Lett.  {\bf 108}, 127205 (2012).

\bibitem{alebrand2012a}S. Alebrand, A. Hassdenteufel, D.  Steil,
M. Cinchetti, and M. Aeschlimann, {Interplay of heating and helicity
in all-optical magnetization switching}, Phys. Rev. B {\bf 85},
092401 (2012).

\bibitem{alebrand2012}S. Alebrand, M. Gottwald, M. Hehn, D. Steil,
M. Cinchetti, D. Lacour, E. E. Fullerton, M. Aeschlimann, and
S. Mangin, {Light-induced magnetization reversal of high-anisotropy
TbCo alloy films}, Appl. Phys. Lett. {\bf 101}, 162408 (2012).

\bibitem{hassdenteufel2013} A. Hassdenteufel, B. Hebler, C.  Schubert,
A. Liebig, M. Teich, M. Helm, M.  Aeschlimann, M. Albrecht, and
R. Bratschitsch, {Thermally assisted all-optical helicity dependent
magnetic switching in amorphous Fe$_{100-x}$Tb$_x$ alloy films},
Adv. Mater. {\bf 25}, 3122 (2013).

\bibitem{mangin2014}S. Mangin, M. Gottwald, C-H. Lambert, D. Steil,
V. Uhlir, L. Pang, M. Hehn, S. Alebrand, M. Cinchetti,
G. Malinowski, Y. Fainman, M. Aeschlimann, and E. E. Fullerton,
{Engineered materials for all-optical helicity-dependent magnetic
switching}, Nat. Mater. {\bf 13}, 286 (2014).

\bibitem{schubert2014a}C. Schubert, A. Hassdenteufel, P. Matthes,
J. Schmidt, M. Helm, R. Bratschitsch, and M. Albrecht, All-optical
helicity dependent magnetic switching in an artificial zero moment
magnet, Appl. Phys. Lett. {\bf 104}, 082406 (2014).

\bibitem{hassdenteufel2015}A. Hassdenteufel, J. Schmidt, C. Schubert,
B. Hebler, M. Helm, M. Albrecht, and R. Bratschitsch, Low-remanence
criterion for helicity-dependent all-optical magnetic switching in
ferrimagnets, Phys. Rev. B {\bf 91}, 104431 (2015).

\bibitem{hassdenteufel2014}A. Hassdenteufel, C. Schubert, B.  Hebler,
H. Schultheiss, J. Fassbender, M. Albrecht, and R.  Bratschitsch,
All-optical helicity dependent magnetic switching in Tb-Fe thin
films with a MHz laser oscillator, Opt.  Express {\bf 22}, 10017
(2014).

\bibitem{gierster2015}L. Gierster, A. A. \"Unal, L. Pape, F. Radu, and
F. Kronast, Laser induced magnetization switching in a TbFeCo
ferrimagnetic thin film: Discerning the impact of dipolar fields,
laser heating and laser helicity by XPEEM, Ultramicroscopy {\bf
159}, 508 (2015).

\bibitem{elhadri2016a}M. S. El Hadri, M. Hehn, P. Pirro,
C.-H. Lambert, G. Malinowski, E. E. Fullerton, and S.  Mangin,
{Domain size criterion for the observation of all-optical
helicity-dependent switching in magnetic thin films}, Phys. Rev. B
{\bf 94}, 064419 (2016).

\bibitem{hebler2016}B. Hebler, A.  Hassdenteufel, P.  Reinhardt, H.
Karl, and M. Albrecht, Ferrimagnetic Tb-Fe alloy thin films:
composition and thickness dependence of magnetic properties and
all-optical switching, Frontiers in Materials {\bf 3}, 8 (2016).

\bibitem{lambert2014} C.-H. Lambert, S. Mangin,
B. S. D. Ch. S. Varaprasad, Y. K. Takahashi, M. Hehn, M. Cinchetti,
G. Malinowski, K. Hono, Y. Fainman, M. Aeschlimann, and
E. E. Fullerton, All-optical control of ferromagnetic thin films
and nanostructures, Science {\bf 345}, 1337 (2014).

\bibitem{john2017}R. John, M. Berritta, D. Hinzke, C. M\"uller,
T. Santos, H. Ulrichs, P. Nieves, J. Walowski, R. Mondal,
O. Chubykalo-Fesenko, J. McCord, P. M. Oppeneer, U. Nowak, and
M. M\"unzenberg, {Magnetisation switching of FePt nanoparticle
recording medium by femtosecond laser pulses}, Sci. Rep. {\bf 7},
4114 (2017).

\bibitem{iihama2018}S. Iihama, Y. Xu, M. Deb, G. Malinowski, M. Hehn,
J. Gorchon, E. E. Fullerton, and S. Mangin, {Single-shot multi-level
all-optical magnetization switching mediated by spin-polarized hot
electron transport}, arXiv 1805.02432.

\bibitem{guyader2015} L. Le Guyader, M. Savoini, S. El Moussaoui,
M. Buzzi, A. Tsukamoto, A. Itoh, A. Kirilyuk, T. Rasing,
A.V. Kimel, and F. Nolting, Nanoscale sub-100 picosecond
all-optical magnetization switching in GdFeCo microstructures,
Nat. Comm. {\bf 6}, 5839 (2015).

\bibitem{ren2017} Y. H.  Ren, W. Lai, Z. Cevher, Y. Gong, and
G. P. Zhang, Experimental demonstration of 55-fs spin canting in
photoexcited iron nanoarrays, Appl. Phys. Lett. {\bf 110}, 082404
(2017).

\bibitem{epl15}G. P. Zhang, Y. H. Bai and T. F. George, {A new and
simple model for magneto-optics uncovers an unexpected spin
switching}, EPL {\bf 112},  27001 (2015).

\bibitem{epl16}G. P. Zhang, Y. H. Bai, and T. F. George, {Switching
ferromagnetic spins by an ultrafast laser pulse: Emergence of giant
optical spin-orbit torque}, EPL {\bf 115}, 57003 (2016).

\bibitem{jpcm17b}G. P. Zhang, Y. H. Bai and T. F. George, {Is
perpendicular magnetic anisotropy essential to all-optical ultrafast
spin reversal in ferromagnets?}  J. Phys.: Condensed Matter {\bf
29}, 425801 (2017).

\bibitem{patrick2018}C. E. Patrick, S. Kumar, K. G\"otze,
M. J. Pearce, J. Singleton, G. Rowlands, G. Balakrishnan,
M. R. Lees, P. A. Goddard, and J. B. Staunton, Field-induced canting
of magnetic moments in GdCo5 at finite temperature: first-principles
calculations and high-field measurements, J. Phys.: Condens. Matter
{\bf 30}, 32LT01 (2018).

\bibitem{elhadri2016}M. S. El Hadri, P. Pirro, C.-H. Lambert,
N. Bergeard, S. Petit-Watelot, M. Hehn, G. Malinowski,
F. Montaigne, Y. Quessab, R. Medapalli, E. E. Fullerton, and
S. Mangin, {Electrical characterization of all-optical
helicity-dependent switching in ferromagnetic Hall crosses},
Appl. Phys. Lett. {\bf 108}, 092405 (2016).

\bibitem{albrecht2018}M. Albrecht, private communication.

\bibitem{yuan2003}S. J. Yuan, L. Sun, H. Sang, J. Du, and S. M. Zhou,
Interfacial effects on magnetic relaxation in Co/Pt multilayers,
Phys. Rev. B {\bf 68}, 134443 (2003).

\bibitem{arx2016}G. P. Zhang, Y. H. Bai, and T. F. George,
arXiv:1609.05855 (2016).

\bibitem{gridnev2018} V. N. Gridnev, Ferromagneticlike states and
all-optical magnetization switching in ferrimagnets, Phys. Rev. B
{\bf 98}, 014427 (2018).

\bibitem{szpunar1977}B. Szpunar and B. Kozarzewski, The application of
CPA to calcuilations of the mean magnetic moment in the
Gd$_{1-x}$Ni$_x$, Gd$_{1-x}$Fe$_x$, Gd$_{1-x}$Co$_x$ and
Y$_{1-x}$Co$_x$ intermetallic compounds, Phys. Stat. Sol. (b) {\bf
82}, 205 (1977).

\bibitem{kurz2002}Ph. Kurz, G. Bihlmayer, and S. Bl\"ugel, {Magnetism and
electronic structure of hcp Gd and the Gd(0001) surface}, J. Phys.:
Condens. Matter {\bf 14}, 6353 (2002).

\bibitem{mentink2012}J. H. Mentink, J. Hellsvik, D. V. Afanasiev,
B. A. Ivanov, A. Kirilyuk, A. V. Kimel, O. Eriksson,
M. I. Katsnelson, and Th. Rasing, Ultrafast Spin Dynamics in
Multisublattice Magnets, Phys. Rev. Lett.  {\bf 108}, 057202 (2012).

\bibitem{prb17a} G. P. Zhang, Z. Babyak, Y. Xue, Y. B. Bai and
T. F. George, {First-principles and model simulation of all-optical
spin reversal}, Phys. Rev. B {\bf 96}, 134407 (2017).

\bibitem{figueroa2014} A. I. Figueroa, J. Bartolom$\rm\acute{e}$,
L. M. García, F. Bartolomé, O. Bunau, J. Stankiewicz, L. Ruiz,
J. M. González-Calbet, F. Petroff, C. Deranlot, S. Pascarelli,
P. Bencok, N. B. Brookes, F. Wilhelm, A. Smekhova, and A. Rogalev,
Structural and magnetic properties of granular Co-Pt multilayers
with perpendicular magnetic anisotropy, Phys. Rev. B {\bf 90},
174421 (2014).

\bibitem{jpcm16}G. P. Zhang, Y. H. Bai, and T. F. George, {Ultrafast
reduction of exchange splitting in ferromagnetic nickel}, J. Phys.:
Condens. Mat. {\bf 28}, 236004 (2016).

\bibitem{pershan1966} P. S. Pershan, J. P. van der Ziei, and
L. D. Malmstrom, Theoretical discussion of the inverse Faraday
effect, Raman scattering, and related phenomena, Phys. Rev. {\bf
143}, 143 (1966).

\bibitem{ren1984} Y. R. Shen, {\it The Principles of Nonlinear
Optics},  (John Wiley \& Sons, Inc., Hoboken, New Jersey, 1984).

\bibitem{jang2016}H. Jang, B. Y. Kang, B. K. Cho, M. Hashimoto, D. Lu,
C. A. Burns, C.-C. Kao, and J.-S. Lee, {Observation of orbital order
in the half-filled $4f$ Gd compound}, Phys. Rev. Lett. {\bf 117},
216404 (2016).

\bibitem{yang2017}Y. Yang, B. Wilson, J. Gorchon, C.-H. Lambert,
S. Salahuddin, and J. Bokor, Ultrafast magnetization reversal by
picosecond electrical pulses. Sci. Adv. {\bf 3}, e1603117 (2017).

\bibitem{baek2018}S.-H. C. Baek, V. P. Amin, Y.-W. Oh, G. Go,
S.-J. Lee, G.-H. Lee, K.-J. Kim, M. D. Stiles, B.-G. Park, and
K.-J. Lee, Spin currents and spin-orbit torques in ferromagnetic
trilayers, Nat. Mat. {\bf 17}, 509 (2018).

\bibitem{brahimi2016}S. Brahimi, H. Bouzar, and S. Lounis, Giant
perpendicular magnetic anisotropy energies in CoPt thin films:
impact of reduced dimensionality and imperfections, J. Phys.:
Condens. Matter {\bf 28}, 496002 (2016).

\bibitem{pham2016} T. H. Pham, J. Vogel, J. Sampaio, M. Vanatka,
J.-C. Rojas-Sannchez, M. Bonfim, D. S. Chaves, F. Choueikani,
P. Ohresser, E. Otero, A. Thiaville, and S. Pizzini, Very large
domain wall velocities in Pt/Co/GdO$_x$ and Pt/Co/Gd trilayers with
Dzyaloshinskii-Moriya interaction, EPL {\bf 113}, 6700 (2016).

\bibitem{liu2016} Z. Y.  Liu, Y. K. Lei and G. F. Wang,
First-principles computation of surface segregation in L1$_0$ CoPt
magnetic nanoparticles, J. Phys.: Condens. Matter {\bf 28}, 266002
(2016).

\bibitem{liu2017} Z. Y. Liu and G. F.  Wang, Surface magnetism of
L1$_0$ CoPt alloy: First principles predictions, J. Phys.:
Condens. Matter {\bf 29}, 355801 (2017).

\bibitem{zimmermann2018} B.  Zimmermann, W. Legrand, N. Reyren,
V. Cros, S. Bl\"uugel, and Albert Fert, Dzyaloshinskii-Moriya
interaction at disordered interfaces from ab initio theory:
robustness against intermixing and tunability through dusting,
arXiv: 1808.04680v1 [cond-mat.mtrl-sci] 14 Aug (2018).

\bibitem{chen2017}J.-Y. Chen, L. He, J.-P. Wang, and M. Li,
All-optical switching of magnetic tunnel junctions with single
subpicosecond laser pulses, Phys. Rev. Appl. {\bf7}, 021001 (2017).

\bibitem{davies2018}C. S. Davies, J. Janus¡onis, A. V. Kimel,
A. Kirilyuk, A. Tsukamoto, Th. Rasing, and R. I. Tobey, Towards
massively parallelized all-optical magnetic recording, J.
Appl. Phys. {\bf 123}, 213904 (2018).

\end{thebibliography}
\end{document}